\documentclass[aps, pra, twocolumn, superscriptaddress]{revtex4-2}

\usepackage[pdftex]{graphicx}
\usepackage{amsmath}
\usepackage{physics}
\usepackage{nicefrac}
\usepackage{siunitx}
\usepackage{bbold}
\usepackage{braket}
\usepackage{ragged2e}
\usepackage{lipsum}
\usepackage{multirow}
\setcitestyle{numbers}
\usepackage{subcaption}
\usepackage{layouts}

\usepackage[dvipsnames]{xcolor}
\usepackage{hyperref}
\hypersetup{
  colorlinks   = true, 
  urlcolor     = MidnightBlue, 
  linkcolor    = MidnightBlue, 
  citecolor   = Maroon 
}

\renewcommand{\arraystretch}{2.0}

\begin{document}

\title{Experimental fault-tolerant code switching}

\author{Ivan Pogorelov}
\altaffiliation{These authors contributed equally}
\affiliation{Universit\"{a}t Innsbruck, Institut f\"{u}r Experimentalphysik, Innsbruck, Austria}

\author{Friederike Butt}
\altaffiliation{These authors contributed equally}
\affiliation{Institute for Quantum Information, RWTH Aachen University, Aachen, Germany}
\affiliation{Institute for Theoretical Nanoelectronics (PGI-2), Forschungszentrum J\"{u}lich, J\"{u}lich, Germany}
    
\author{Lukas Postler}
\affiliation{Universit\"{a}t Innsbruck, Institut f\"{u}r Experimentalphysik, Innsbruck, Austria}

\author{\mbox{Christian D. Marciniak}}
\affiliation{Universit\"{a}t Innsbruck, Institut f\"{u}r Experimentalphysik, Innsbruck, Austria}

\author{Philipp Schindler}
\affiliation{Universit\"{a}t Innsbruck, Institut f\"{u}r Experimentalphysik, Innsbruck, Austria}

\author{Markus M\"{u}ller}
\affiliation{Institute for Quantum Information, RWTH Aachen University, Aachen, Germany}
\affiliation{Institute for Theoretical Nanoelectronics (PGI-2), Forschungszentrum J\"{u}lich, J\"{u}lich, Germany}

\author{Thomas Monz}
    \altaffiliation[Also at ]{Alpine Quantum Technologies GmbH, Innsbruck, Austria}
\email[Email to ]{thomas.monz@uibk.ac.at}
\affiliation{Universit\"{a}t Innsbruck, Institut f\"{u}r Experimentalphysik, Innsbruck, Austria}

\date{\today}

\begin{abstract}
Quantum error correction is a crucial tool for mitigating hardware errors in quantum computers by encoding logical information into multiple physical qubits. However, no single error-correcting code allows for an intrinsically fault-tolerant implementation of all the gates needed for universal quantum computing~\cite{eastin2009restrictions, kitaev1997quantum, solovay1995lie}.  
One way to tackle this problem is to switch between two suitable error-correcting codes, while preserving the encoded logical information, which in combination give access to a fault-tolerant universal gate set~\cite{anderson2014fault, bombin2016dimensional, kubica2015universal}.
In this work, we present the first experimental implementation of fault-tolerant code switching between two codes. One is the seven-qubit color code~\cite{steane1996multiple}, which features fault-tolerant CNOT and $H$ quantum gates, while the other one, the 10-qubit code~\cite{vasmer2022morphing}, allows for a fault-tolerant $T$-gate implementation. Together they form a complementary universal gate set. 
Building on essential code switching building blocks, we construct logical circuits and prepare 12 different logical states which are not accessible natively in a fault-tolerant way within a single code. 
Finally, we use code switching to entangle two logical qubits employing the full universal gate set in a single logical quantum circuit. 
Our results experimentally open up a new route towards deterministic control over logical qubits with low auxiliary qubit overhead, not relying on the probabilistic preparation of resource states.

\end{abstract}

\maketitle

\section{Introduction}

Quantum computers offer the prospect of performing certain computational tasks more efficiently than any known classical algorithm~\cite{shor1994algorithms, grover1997quantum}. 
However, the accuracy of quantum computations is limited by noise, arising from the interaction of qubits with their environment~\cite{preskill2018quantum}. Quantum error correction (QEC) addresses this challenge by encoding quantum information across multiple physical qubits, thereby adding redundancy and allowing errors to be localized and corrected without destroying the encoded information~\cite{gottesman1997stabilizer, Nielsen_and_Chuang}. 
QEC is performed by physical operations which are themselves faulty. Operations on these error-corrected logical qubits therefore have to be performed without proliferating local errors uncontrollably across the encoded qubit. 
This can be achieved by means of fault-tolerant (FT) circuit designs~\cite{kitaev1997quantum, aliferis2005quantum, knill1998resilient, aharonov1997fault} for example by using very resource-efficient transversal gates. Transversal gate operations perform logical operations by qubitwise application of the physical operations.
Generally, it is not possible to implement an arbitrary logical operation transversally in a given QEC code. Computations on qubits can be approximated to arbitrary precision using only a finite set of gates forming a so-called universal gate set~\cite{solovay1995lie, Nielsen_and_Chuang}, as for example the gate set consisting of the Hadamard-gate ($H$), $T$-gate, and two-qubit entangling CNOT-gate would allow for universal quantum computing. 
However, at least one of these gates cannot be transversal for encoded qubits ~\cite{eastin2009restrictions}.
Consequently, at least one logical gate must be implemented through alternative methods to achieve universal FT computation, which often implies a substantial resource overhead~\cite{eastin2009restrictions}.

\begin{figure*}[!htb]
	\centering
	\includegraphics[width=120mm]{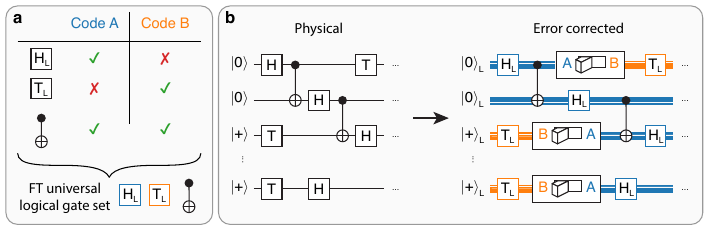}
	\caption{\justifying \textbf{Quantum algorithms with universal gate set. } (a) Two codes with complementary sets of FT gates in combination amount to a complete FT universal logical gate set \{$H_L, T_L, \mathrm{CNOT}_L$\}. (b) An algorithm running on physical qubits (left) can be run fault-tolerantly on logical qubits (right) by switching between two codes. If, for example, the FT $T_L$ gate is only available in code B (orange), one has to switch to this code before applying the respective gate operation. }
\label{fig:universal_gate_set_motivation}
\end{figure*}

Two well-known methods to complete a universal gate set are magic state injection and code switching. 
With magic state injection, it is possible to implement a non-Clifford gate by preparing a magic resource state fault-tolerantly on a logical auxiliary qubit~\cite{goto2016minimizing, chamberland2019fault} and injecting this auxiliary state onto the encoded data qubit~\cite{bravyi2005universal}. 
Magic states have been prepared on superconducting architectures~\cite{gupta2024encoding}, as well as on ion trap quantum processors~\cite{egan2021fault}, and a full universal gate set has recently been realized on an ion-trap quantum processor~\cite{postler2022demonstration}. 
However, producing high-fidelity magic states poses a significant challenge and presents a large overhead~\cite{gidney2019efficient}.
As an alternative method, switching between two codes with complementary sets of transversal gates also enables the implementation of a full universal gate set~\cite{kubica2015universal, bombin2016dimensional, anderson2014fault}, as illustrated in Fig.~\ref{fig:universal_gate_set_motivation}. Here, the key idea is to transfer encoded information from one encoding to another by measuring a set of operators and applying local Pauli operations. These are chosen in a way that fixes the state into the desired codespace but also preserves the encoded information. 
An experimental demonstration of this route to fault-tolerance has so far not been achieved.

In previous theoretical work we developed FT code switching protocols utilizing the concept of flag qubits, as introduced by~\cite{chao2018quantum, chamberland2018flag, goto2016minimizing}, 
and found schemes that reach a level of performance competitive with magic state injection~\cite{butt2023fault}. 
In this work we present the first experimental implementation of such FT code switching protocols. We characterize the performance of essential building blocks for switching between the seven-qubit color code, encoding a single logical qubit and capable of correcting a single error, and an error-detecting 10-qubit code~\cite{vasmer2022morphing} on an ion-trap quantum processor. We prepare 12 different states on the seven-qubit color code, which can not be obtained transversally using a single error-correcting code. We furthermore extend these protocols to entangling logical operations and implement minimal logical algorithms using all gates of a universal gate set \{$H, T, \mathrm{CNOT}$\}.

\section{FT code switching procedure}\label{sec:theory}

\begin{figure*}[!htb]
	\centering
	\includegraphics[width=180mm]{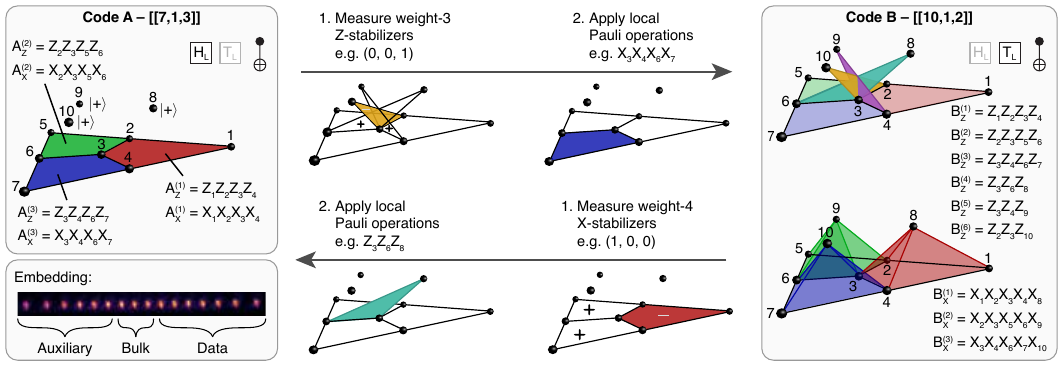}
	\caption{\justifying \textbf{Switching between $[[7, 1, 3]]$ and $[[10, 1, 2]]$. } A valid codestate of the $[[7, 1, 3]]$ code (left) is a +1-eigenstate of the $X$- and $Z$-stabilizers $A^{(i)}_X$ and $A^{(i)}_Z$ which are defined symmetrically on the weight-4 plaquettes (red, green, blue). In contrast to that, the $X$- and $Z$-stabilizers of the $[[10, 1, 2]]$ code (right) differ: $Z$-stabilizers are defined on the three weight-4 plaquettes in the triangular plane (red, green blue), as well as on the three weight-3 plaquettes connecting the triangular plane with the bulk qubits 8, 9, and 10 (orange, turquoise, purple). The $X$-stabilizers of the $[[10, 1, 2]]$ code have support on the weight-5 pyramids (red, green, blue). To switch from the $[[7, 1, 3]]$ code to the $[[10, 1, 2]]$ code, we measure the three weight-3  $Z$-stabilizers which connect the encoded $[[7, 1, 3]]$ on qubits 1--7 with the bulk qubits 8, 9 and 10. Based on the obtained switching syndrome, a combination of the weight-4 plaquettes $(A^{(1)}_X, A^{(2)}_X, A^{(3)}_X)$ is applied, which fixes the logical state into the desired codespace while preserving the encoded information. The inverse switching direction is implemented analogously: first, the three weight-4 $X$-stabilizers are measured, followed by the application of a combination of $(B^{(4)}_Z, B^{(5)}_Z, B^{(6)}_Z)$. }
\label{fig:overview_switching_ion_trap}
\end{figure*}

The seven-qubit color code encodes $k = 1$ logical qubit in $n = 7$ physical qubits and has a code distance $d = 3$, allowing the correction of any single error~\cite{steane1996multiple, bombin2006topological}. A valid logical state is the simultaneous +1-eigenstate of the stabilizer generators $A_{\sigma}^{(i)}$, which are shown in Fig.~\ref{fig:overview_switching_ion_trap} on the left and given explicitly in App.~\ref{app:EC_Codes}. 
The logical Pauli operators of this code correspond to applying Pauli $X$- and $Z$- operations to all seven qubits $X_L=X^{\otimes 7}$ and $Z_L=Z^{\otimes 7}$. A logical Hadamard gate $H_L$ can be implemented transversally by bitwise application of single-qubit Hadamard-gates and, similarly, the phase gate $S_L$ can be realized in a transversal manner by applying single-qubit $S^{\dag}$-gates to all qubits. 
As introduced in~\cite{vasmer2022morphing}, the $[[10, 1, 2]]$ code encodes a single logical qubit in 10 physical qubits and, with distance $d = 2$, can \textit{detect} any single error. 
The logical qubit of the $[[10, 1, 2]]$ code is defined by the stabilizer generators $B_{\sigma}^{(i)}$, shown in Fig.~\ref{fig:overview_switching_ion_trap} on the right and given in App.~\ref{app:EC_Codes}, and the logical Pauli-operators coincide with those of the seven-qubit color code. 
$X$- and $Z$-stabilizer generators are not defined on the same support, meaning that the $[[10, 1, 2]]$ code is not self-dual, and consequently does not have a transversal Hadamard-gate. 
It is generated from the larger 15-qubit tetrahedral code, inheriting its FT non-Clifford $T$-gate, which can be implemented with~\cite{vasmer2022morphing}
\begin{align}
    T_L = T_1 T^{\dag}_2 T_3 T^{\dag}_4 T_5 T^{\dag}_6 T_7 \, \mathrm{CCZ}_{8, 9, 10}, 
\end{align}
which includes the controlled-controlled-Z-gate on qubits 8, 9, and 10. This implementation of the logical $T_L$-gate is not transversal anymore but FT in the sense that all possible errors resulting from a single fault are still detectable. The $[[10, 1, 2]]$ code is the smallest known code that has a FT $T$-gate~\cite{vasmer2022morphing}. 

We implement a FT universal gate set by switching between the seven-qubit color code $[[7, 1, 3]]$ and the $[[10, 1, 2]]$ code~\cite{steane1996multiple, bombin2006topological, vasmer2022morphing}. 
We can transfer encoded information between these two codes by first measuring those stabilizers of the target code which differ from the stabilizers of the initial code. This measurement randomly projects the logical state onto a $\pm$1-eigenstate of the measured stabilizers. In a second step, we force the state into the +1-eigenstate of the measured stabilizers without changing the logical state~\cite{poulin2005stabilizer, anderson2014fault, kubica2015universal} by applying a combination of Pauli-generators. Here, these generators directly correspond to stabilizer operators of the initial code and, therefore, do not affect the logical state. 

Specifically, we measure the three $Z$-stabilizers $(B^{(4)}_Z, B^{(5)}_Z, B^{(6)}_Z)$, shown in Fig.~\ref{fig:overview_switching_ion_trap} to switch from $[[7, 1, 3]]$ to $[[10, 1, 2]]$. 
Then, we apply a combination of the weight-4 $X$-generators of the $[[7, 1, 3]]$ code $(A^{(1)}_X, A^{(2)}_X, A^{(3)}_X)$. For example, if the random projection onto the $Z$-stabilizers yields $(0, 0, 1)$, where 0 corresponds to a +1-eigenvalue and 1 to a -1-eigenvalue of the measured operator, we would apply \mbox{$A^{(3)}_X = X_3 X_4 X_6 X_7$}. $A^{(3)}_X$ overlaps at an even number of sites with the first and second $Z$-stabilizer and only at a single site with $B^{(6)}_X$ and, therefore, this operation fixes the state into the +1-eigenstate of all $Z$-stabilizers $(B^{(4)}_Z, B^{(5)}_Z, B^{(6)}_Z)$ of the ten-qubit code. 
For switching in the inverse direction from $[[10, 1, 2]]$ to $[[7, 1, 3]]$ , we employ the same scheme but interchange the sets of stabilizers: we measure $(A^{(1)}_X, A^{(2)}_X, A^{(3)}_X)$ and apply a combination of $(B^{(4)}_Z, B^{(5)}_Z, B^{(6)}_Z)$. The lookup table with possible measurement outcomes and switching operations for both directions is summarized in App. Tab.~\ref{tab:lookuptable_switching}. 

To achieve fault tolerance, we repeat measurements to correct for single measurement faults. We use flag-qubits for stabilizer measurements in order to prevent faults on auxiliary qubits from propagating uncontrollably~\cite{chao2018quantum, chamberland2018flag, goto2016minimizing, hilder2022fault, butt2023fault}, and we perform additional stabilizer measurements to detect potentially dangerous errors on data qubits. Whenever a potentially dangerous error is detected, the corresponding run is discarded. The full FT protocols for switching in both directions are discussed and summarized in App.~\ref{app:EC_Codes}.

\section{Experimental setup}\label{sec:experimental_setup}

The experiment was performed with a 16-ion chain of $^{40}$\textrm{Ca}$^+$ ions trapped in a linear Paul trap. We utilize an optical qubit encoded in $\ket{0}=\ket{4 ^2\textrm{S}_{1/2}, m_J = -1/2}$ and $\ket{1}=\ket{3 ^2\textrm{D}_{5/2}, m_J = -1/2}$ Zeeman sub-levels. An optical addressing system for \SI{729}{\nano\meter} laser light driving the qubit transition allows for individual qubit control. In addition, ion-ion interaction through common motional modes of the trap provides all-to-all connectivity for two-qubit gates based on the M{\o}lmer-S{\o}rensen (MS) interaction~\cite{sorensen2000entanglement}. A more detailed description of the experimental setup is given in~\cite{pogorelov2021compact, postler2022demonstration, heussen2023strategies}. 

The setup features detection of the ion chain using the electron shelving technique such that only a desired subset of qubits is detected while the state of the other qubits is preserved. Subsequent recooling of the ion chain and reinitialization of the detected qubits allow for applying high-fidelity gates after the detection procedure. This procedure is used here for mid-circuit measurements of the auxiliary qubits for flag-based stabilizer readout. A comprehensive overview of the procedure can be found in~\cite{postler2023demonstration}.
The error rates for the basic experimental qubit manipulation operations are given in App. Tab.~\ref{tab:error_rates_simulation}.

\section{Results}
\label{sec:results}

\begin{figure}[t]
	\centering
	\includegraphics[width=89mm]{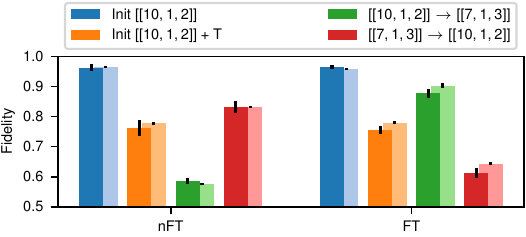}
	\caption{\justifying \textbf{Fidelities for code switching building blocks.} Logical process fidelities for nFT (left) and FT (right) protocols for the initialization of $[[10, 1, 2]]$ logical states (blue), a logical $T$-gate (orange), switching from $[[10, 1, 2]]$ to $[[7, 1, 3]]$ (green) and in the inverse direction (red). 
    The values obtained in the experiment/simulation are depicted in darker/lighter colors, respectively. The error bars show standard deviations, determined as discussed in App.~\ref{tomo_details}, and explicit values are summarized in App. Tab.~\ref{tab:fidelities_building_blocks}. }
	\label{fig:blocks}
\end{figure}

\begin{figure*}[tb]
	\centering
	\includegraphics[width=120mm]{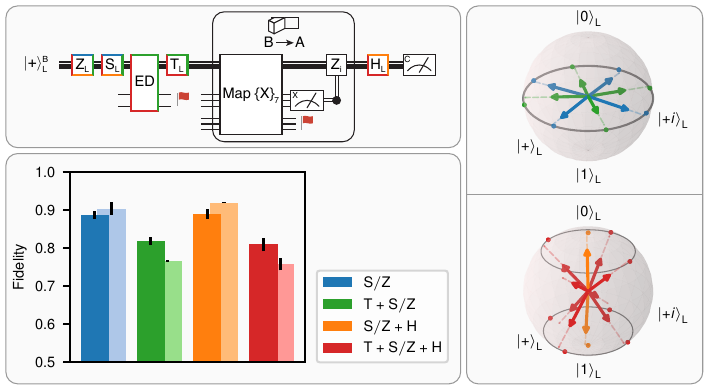}
	\caption{\justifying \textbf{Spanning logical states on the Bloch sphere. }
 (a) Circuit used in the experiment to produce different logical states on the Bloch sphere. The colored lines around gates indicate if the gate was executed to prepare the state of the respective color, e.g. the $T$-gate was only performed when producing 'green' or 'red' states.
 First, a logical $|+\rangle_L$ is initialized for the $[[10, 1, 2]]$ code, followed by operations from the gate set $\{Z, S, T\}$. The $T$-gate is made FT by including an additional error-detection block before applying the $T$-gate. We then switch to the $[[7, 1, 3]]$ code by measuring three weight-4 $X$-stabilizers and applying a $Z$-type switching operation based on the obtained measurement outcomes. Finally, we can perform $\pi/2$ rotations about the $X$- or $Y$-axis, which correspond to a transversal logical $H$-gate. 
 (b) Bloch vectors for all states that are reachable with the different gate combinations of the protocol shown in (a). These include the four cardinal states on the equator of the Bloch sphere (blue), as well as those four states which are rotated by $\ang{45}$ about the $Z$-axis (green). The north and south poles (orange) are reached by rotating about the $Y$-axis. By applying a combination of all operations, i.e. the initial (non-)Clifford gates $\{S, Z, T\}$ and the final rotation, we can reach various points on $45^\mathrm{th}$ parallels (gray circles) of the Bloch sphere (red).
(c) State fidelities for different logical states with the circuit in (a). These are averaged over groups of states with the same number of physical operations. For example, the blue bar corresponds to the fidelity averaged over the blue Bloch vectors in (b). Experimental results are shown in darker and simulated results in lighter colors. The error bars show standard deviations, determined as discussed in App.~\ref{tomo_details}, and the values are summarized in App. Tab.~\ref{tab:trivial_syndrome}.
}   
	\label{fig:bloch}
\end{figure*}

\begin{figure*}[htb]
	\centering
	\includegraphics[width=120mm]{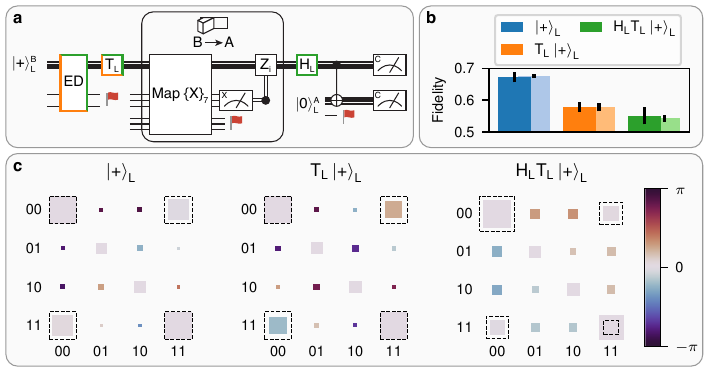}
	\caption{\justifying \textbf{Entangled states with non-Clifford gates by means of code switching.} 
 (a) Protocol for the preparation of entangled logical qubits. 
 A logical Bell-state $\frac{1}{\sqrt{2}}(|00\rangle_L + |11\rangle_L)$ is prepared by encoding a qubit on the $[[10, 1, 2]]$ code and FT switching to $[[7, 1, 3]]$. A second qubit is then initialized in $|0\rangle_L$ on the seven-qubit color code and coupled to the initial logical qubit by applying a transversal CNOT-gate. $\frac{1}{\sqrt{2}}(|00\rangle_L + e^{i\frac{\pi}{4}}|11\rangle_L)$ and $\cos{\frac{\pi}{8}} |00\rangle_L + \sin{\frac{\pi}{8}}|11\rangle_L$ are prepared by including the FT $T$-gate on the initial 10-qubit code and the $\pi/2$ rotations about the $X$-axis on the $[[7, 1, 3]]$ code, which corresponds to a transversal logical $H$-gate.  Colored lines around gates indicate which gates were used to prepare the state of the corresponding color. 
 (b) Fidelities for the entangled logical states which can be prepared with the different gate combinations of protocol (a). Experiment/simulation results are shown in darker/lighter colors. The error bars show standard deviations, determined as discussed in App.~\ref{tomo_details}, and the values are summarized in App. Tab.~\ref{tab:trivial_syndrome}.
 (c) Logical state tomography result for the three entangled states. The dashed black boxes indicate the ideal values which correspond to the fault-free case. }
	\label{fig:cnot}
\end{figure*}

We characterize each building block essential for implementing the universal gate set based on code switching: the initialization of $[[10, 1, 2]]$ logical states, the logical $T$-gate, and switching back and forth between the $[[10, 1, 2]]$ and the $[[7, 1, 3]]$ codes. We perform logical quantum process tomography, as specified in App.~\ref{app:experiment}, for all building blocks and compare the results to the respective ideal process. All experimental results are accompanied by numerical simulations using a multi-parameter noise model described in App.~\ref{app:numerics} which is using experimental error rates given in App. Tab.~\ref{tab:error_rates_simulation}. The resulting logical process fidelities are shown in Fig.~\ref{fig:blocks}. All obtained fidelities and acceptance rates are given in App. Tab.~\ref{tab:fidelities_building_blocks}. 

We find that the fidelities for the initialization in the nFT case are similar to those in the FT case, since the FT overhead for an additional verification is small and errors do not propagate in a dangerous way for the nFT protocol. Only if further operations are performed on the logical state afterward, previously correctable errors can become uncorrectable even for transversal operations due to different code distances for $X$- and $Z$-errors. 
For the $T$-gate on the $[[10, 1, 2]]$ code we observe a similar feature as the fidelity does not change significantly when employing the FT scheme. The FT $T$-gate on the $[[10, 1, 2]]$ code includes one additional minimal error detection (ED) block before applying the logical $T$-gate, which is reflected in a decreasing acceptance rate: 81\% are accepted for the nFT protocol while 51\% are accepted for the FT version, as summarized in App. Tab.~\ref{tab:fidelities_building_blocks}.  
We can identify a clear advantage of the FT scheme over the nFT scheme for switching from $[[10, 1, 2]]$ to $[[7, 1, 3]]$, as we can achieve fault tolerance with only a small increase of the two-qubit gate count as discussed in Sec.~\ref{sec:theory}. 
However, the FT switching protocol does not provide an advantage over the nFT one for $[[7, 1, 3]] \rightarrow [[10, 1, 2]]$. This is not unexpected due to the complexity of the protocol with a large circuit depth and the current error rates, which are discussed in detail in App.~\ref{app:FT_building_blocks}. 

As a next step, we combine the aforementioned building blocks to prepare a variety of different states fault-tolerantly, which are inaccessible with natively FT gate implementations in the seven-qubit color code. 
The protocol is shown in Fig.~\ref{fig:bloch}\textbf{a}: initially, $\ket{+}_L$ of the $[[10, 1, 2]]$ is prepared followed by the application of FT gates of the $[[10, 1, 2]]$ code which includes any combinations of gates from the set $\{Z, S, T\}$. 
This enables the preparation of the four cardinal states as well as four additional states, which require a $T$-gate, on the equator of the Bloch sphere, as shown in Fig.~\ref{fig:bloch}\textbf{b}.
After switching to the seven-qubit color code, we can lastly apply an $H$-gate. This allows us to prepare logical states on the 45th parallels of the Bloch sphere, which are not reachable with a single code with a natively FT implementation. 
We implement this protocol experimentally and numerically and perform logical quantum state tomography for each obtained state. The reconstructed Bloch vectors for all prepared logical states are shown in Fig.~\ref{fig:bloch}\textbf{b}. 
Circuits including the non-Clifford $T$-gate have decreased fidelity relative to purely Clifford circuits. This can be attributed to the additional two-qubit gates required for the ED block before the application of the $T$-gate, and the mid-circuit measurement which does not have to be performed for the Clifford circuits. 
We identify dephasing on idling qubits as a major contribution to the total logical state infidelity and take further measures to improve the fidelities of the states, prepared with the protocol in Fig.~\ref{fig:bloch}, which is discussed further in App.~\ref{app:error_budget}. 
We can effectively increase the fidelities by up to 0.08 for Clifford states and up to 0.04 for non-Clifford states by simply reassigning the $X$- and $Z$- stabilizers, which is shown in App. Fig.~\ref{fig:bloch_inverted}. 

We extend the demonstrated set of logical gates by entangling two-qubit operations with the goal of exploring small logical circuits with gates from the universal gate set. We again start in the logical $|+\rangle_L$ of the $[[10, 1, 2]]$ code and may apply a FT $T$-gate before switching to $[[7, 1, 3]]$ as shown in Fig.~\ref{fig:cnot}\textbf{a}. On the target code, we can apply an additional $H$-gate. Afterward, a second logical qubit is prepared on the $[[7, 1, 3]]$ code in $\ket{0}_L$ and the two logical qubits are entangled with a logical CNOT. 
Utilizing the full universal gate set, we prepare three entangled states with this protocol and analyze these states by means of logical quantum state tomography, as shown in Fig.~\ref{fig:cnot}\textbf{b, c} and explicitly given in App.~\ref{tomo_details}.
Again, experimental results and the numerically obtained fidelities agree within one standard deviation. 
All three logical states are entangled with $>99\%$ confidence despite a large circuit depth and circuit complexity, involving up to 61 two-qubit gates and two mid-circuit measurements. 

\section{Discussion}

In our experimental demonstration of FT code switching we operate all gates directly at the logical level - code switching between error-correcting codes does not rely on probabilistic preparation of resource states. 
This fact is important in light of the observation that a deterministic code execution empowered by code switching rather than a probabilistic protocol via state injection can become advantageous over probabilistic execution via state preparation and injection, as overall device capabilities improve.~\cite{heussen2023strategies}.

We find experimentally that fault-tolerant code switching schemes can significantly improve fidelities in certain situations compared to their non-fault-tolerant counterparts. However, this advantage does currently not yet extend to protocols with large circuit depth or many mid-circuit measurements. 
These primary bottlenecks in our implementation fidelity can be mitigated with technical or logical changes. 
In our particular implementation we may benefit mostly from independently-demonstrated technical improvements such as extended coherence times~\cite{harty2014high, ruster2016long, wang2021single} and more robust composite pulse sequences during mid-circuit measurements~\cite{wimperis1989composite,wimperis1994broadband}. 
Furthermore, tailoring QEC protocols to systems with biased noise~\cite{bonilla2021xzzx, xu2023tailored, huang2023tailoring, pal2022relaxation} promises the possibility of reaching higher fidelities for experimental setups with these noise characteristics.  
The present case of $Z$-bias makes selecting a code like the rotated [[10,1,2]]$_X$ which is more resilient to $Z$-noise beneficial, as demonstrated by the improved performance under this exchange of stabilizers.

In the near future code switching can be beneficial in hardware systems where the cost of mid-circuit measurements is low compared to the cost of having additional auxiliary qubits for the magic state injection. Contemporary superconducting quantum computers with limited connectivity~\cite{krinner2022realizing, kelly2015state, zhao2022realization, google2021exponential, satzinger2021realizing, takita2017experimental} and QCCD-based trapped-ion architectures~\cite{pino2021demonstration, ryan2021realization, wan2020ion, hilder2022fault} might fall into this category.  

Here, we have considered the experimental implementation of code switching for the smallest possible instances of the respective classes of codes, namely the $[[10, 1, 2]]$ and the $[[7, 1, 3]]$ code. Future work will include the extension from the error-detecting $[[10, 1, 2]]$ code to the deterministic error-correcting $[[15, 1, 3]]$ code, which has been explored theoretically~\cite{butt2023fault, anderson2014fault}, as well as larger-distance codes, which promise the implementation of a deterministic set of universal gates, without the need for large-scale magic-state factories. 

\section{Acknowledgements}
We gratefully acknowledge support by the European Union’s Horizon Europe research and innovation program under Grant Agreement Number 101114305 (“MILLENION-SGA1” EU Project), the US Army Research Office through Grant Number W911NF-21-1-0007, the European Union’s Horizon Europe research and innovation program under Grant Agreement Number 101046968 (BRISQ), the ERC Starting Grant QCosmo under Grant Number 948893, the ERC Starting Grant QNets through Grant Number 804247, the Austrian Science Fund (FWF Grant-DOI 10.55776/F71) (SFB BeyondC), the Austrian Research Promotion Agency under Contracts Number 896213 (ITAQC) and 897481 (HPQC) supported by the European Union – NextGenerationEU, the Office of the Director of National Intelligence (ODNI), Intelligence Advanced Research Projects Activity (IARPA), under the Entangled Logical Qubits program through Cooperative Agreement Number W911NF-23-2-0216. This research is also part of the Munich Quantum Valley (K-8), which is supported by the Bavarian state government with funds from the Hightech Agenda Bayern Plus.
We further receive support from the IQI GmbH, and by the German ministry of science and education (BMBF) via the VDI within the project IQuAn, and by the Deutsche Forschungsgemeinschaft (DFG, German Research Foundation) under Germany’s Excellence Strategy ‘Cluster of Excellence Matter and Light for Quantum Computing (ML4Q) EXC 2004/1’ 390534769.

The views and conclusions contained in this document are those of the authors and should not be interpreted as representing the official policies, either expressed or implied, of IARPA, the Army Research Office, or the U.S. Government. The U.S. Government is authorized to reproduce and distribute reprints for Government purposes notwithstanding any copyright notation herein.

We acknowledge computing time provided at the NHR Center NHR4CES at RWTH Aachen University (Project No. p0020074). This is funded by the Federal Ministry of Education and Research and the state governments participating on the basis of the resolutions of the GWK for national high-performance computing at universities. 

\subsection*{Data availability}
The data provided in the figures in this article and the circuits used in the experiment are available at \url{https://doi.org/10.5281/zenodo.10812902}. Full measurement data can be provided by the corresponding author upon reasonable request.

\subsection*{Authors contribution:}
I.P. carried out the experiments. I.P., L.P., C.D.M. contributed to the experimental setup. I.P., F.B. analyzed the data. F.B. developed the protocols and underlying quantum circuits, performed the numerical simulations, characterization, and theory modeling. I.P., F.B., and C.D.M. wrote the manuscript, with contributions from all authors. P.S., M.M., and T.M. supervised the project.

\subsection*{Competing interests:}
T.M. is connected to Alpine Quantum Technologies GmbH, a commercially oriented quantum computing company.

\newpage

\bibliography{references}

\begin{thebibliography}{57}%
\makeatletter
\providecommand \@ifxundefined [1]{%
 \@ifx{#1\undefined}
}%
\providecommand \@ifnum [1]{%
 \ifnum #1\expandafter \@firstoftwo
 \else \expandafter \@secondoftwo
 \fi
}%
\providecommand \@ifx [1]{%
 \ifx #1\expandafter \@firstoftwo
 \else \expandafter \@secondoftwo
 \fi
}%
\providecommand \natexlab [1]{#1}%
\providecommand \enquote  [1]{``#1''}%
\providecommand \bibnamefont  [1]{#1}%
\providecommand \bibfnamefont [1]{#1}%
\providecommand \citenamefont [1]{#1}%
\providecommand \href@noop [0]{\@secondoftwo}%
\providecommand \href [0]{\begingroup \@sanitize@url \@href}%
\providecommand \@href[1]{\@@startlink{#1}\@@href}%
\providecommand \@@href[1]{\endgroup#1\@@endlink}%
\providecommand \@sanitize@url [0]{\catcode `\\12\catcode `\$12\catcode `\&12\catcode `\#12\catcode `\^12\catcode `\_12\catcode `\%12\relax}%
\providecommand \@@startlink[1]{}%
\providecommand \@@endlink[0]{}%
\providecommand \url  [0]{\begingroup\@sanitize@url \@url }%
\providecommand \@url [1]{\endgroup\@href {#1}{\urlprefix }}%
\providecommand \urlprefix  [0]{URL }%
\providecommand \Eprint [0]{\href }%
\providecommand \doibase [0]{https://doi.org/}%
\providecommand \selectlanguage [0]{\@gobble}%
\providecommand \bibinfo  [0]{\@secondoftwo}%
\providecommand \bibfield  [0]{\@secondoftwo}%
\providecommand \translation [1]{[#1]}%
\providecommand \BibitemOpen [0]{}%
\providecommand \bibitemStop [0]{}%
\providecommand \bibitemNoStop [0]{.\EOS\space}%
\providecommand \EOS [0]{\spacefactor3000\relax}%
\providecommand \BibitemShut  [1]{\csname bibitem#1\endcsname}%
\let\auto@bib@innerbib\@empty
\bibitem [{\citenamefont {Eastin}\ and\ \citenamefont {Knill}(2009)}]{eastin2009restrictions}%
  \BibitemOpen
  \bibfield  {author} {\bibinfo {author} {\bibfnamefont {B.}~\bibnamefont {Eastin}}\ and\ \bibinfo {author} {\bibfnamefont {E.}~\bibnamefont {Knill}},\ }\bibfield  {title} {\bibinfo {title} {Restrictions on transversal encoded quantum gate sets},\ }\href {https://doi.org/10.1103/PhysRevLett.102.110502} {\bibfield  {journal} {\bibinfo  {journal} {Phys. Rev. Lett.}\ }\textbf {\bibinfo {volume} {102}},\ \bibinfo {pages} {110502} (\bibinfo {year} {2009})}\BibitemShut {NoStop}%
\bibitem [{\citenamefont {Kitaev}(1997)}]{kitaev1997quantum}%
  \BibitemOpen
  \bibfield  {author} {\bibinfo {author} {\bibfnamefont {A.~Y.}\ \bibnamefont {Kitaev}},\ }\bibfield  {title} {\bibinfo {title} {Quantum computations: algorithms and error correction},\ }\href {https://doi.org/10.1070/RM1997v052n06ABEH002155} {\bibfield  {journal} {\bibinfo  {journal} {Russ. Math. Surv.}\ }\textbf {\bibinfo {volume} {52}},\ \bibinfo {pages} {1191} (\bibinfo {year} {1997})}\BibitemShut {NoStop}%
\bibitem [{\citenamefont {Solovay}(1995)}]{solovay1995lie}%
  \BibitemOpen
  \bibfield  {author} {\bibinfo {author} {\bibfnamefont {R.}~\bibnamefont {Solovay}},\ }\href@noop {} {\bibinfo {title} {Lie groups and quantum circuits, 2000}} (\bibinfo {year} {1995})\BibitemShut {NoStop}%
\bibitem [{\citenamefont {Anderson}\ \emph {et~al.}(2014)\citenamefont {Anderson}, \citenamefont {Duclos-Cianci},\ and\ \citenamefont {Poulin}}]{anderson2014fault}%
  \BibitemOpen
  \bibfield  {author} {\bibinfo {author} {\bibfnamefont {J.~T.}\ \bibnamefont {Anderson}}, \bibinfo {author} {\bibfnamefont {G.}~\bibnamefont {Duclos-Cianci}},\ and\ \bibinfo {author} {\bibfnamefont {D.}~\bibnamefont {Poulin}},\ }\bibfield  {title} {\bibinfo {title} {Fault-tolerant conversion between the {S}teane and {R}eed-{M}uller quantum codes},\ }\href {https://doi.org/10.1103/PhysRevLett.113.080501} {\bibfield  {journal} {\bibinfo  {journal} {Phys. Rev. Lett.}\ }\textbf {\bibinfo {volume} {113}},\ \bibinfo {pages} {080501} (\bibinfo {year} {2014})}\BibitemShut {NoStop}%
\bibitem [{\citenamefont {Bomb{\'\i}n}(2016)}]{bombin2016dimensional}%
  \BibitemOpen
  \bibfield  {author} {\bibinfo {author} {\bibfnamefont {H.}~\bibnamefont {Bomb{\'\i}n}},\ }\bibfield  {title} {\bibinfo {title} {Dimensional jump in quantum error correction},\ }\href {https://doi.org/10.1088/1367-2630/18/4/043038} {\bibfield  {journal} {\bibinfo  {journal} {NJP}\ }\textbf {\bibinfo {volume} {18}},\ \bibinfo {pages} {043038} (\bibinfo {year} {2016})}\BibitemShut {NoStop}%
\bibitem [{\citenamefont {Kubica}\ and\ \citenamefont {Beverland}(2015)}]{kubica2015universal}%
  \BibitemOpen
  \bibfield  {author} {\bibinfo {author} {\bibfnamefont {A.}~\bibnamefont {Kubica}}\ and\ \bibinfo {author} {\bibfnamefont {M.~E.}\ \bibnamefont {Beverland}},\ }\bibfield  {title} {\bibinfo {title} {Universal transversal gates with color codes: A simplified approach},\ }\href {https://doi.org/10.1103/PhysRevA.91.032330} {\bibfield  {journal} {\bibinfo  {journal} {Phys. Rev. A}\ }\textbf {\bibinfo {volume} {91}},\ \bibinfo {pages} {032330} (\bibinfo {year} {2015})}\BibitemShut {NoStop}%
\bibitem [{\citenamefont {Steane}(1996)}]{steane1996multiple}%
  \BibitemOpen
  \bibfield  {author} {\bibinfo {author} {\bibfnamefont {A.}~\bibnamefont {Steane}},\ }\bibfield  {title} {\bibinfo {title} {Multiple-particle interference and quantum error correction},\ }\href {https://doi.org/10.1098/rspa.1996.0136} {\bibfield  {journal} {\bibinfo  {journal} {Proc. R. Soc. London, Ser. A}\ }\textbf {\bibinfo {volume} {452}},\ \bibinfo {pages} {2551} (\bibinfo {year} {1996})}\BibitemShut {NoStop}%
\bibitem [{\citenamefont {Vasmer}\ and\ \citenamefont {Kubica}(2022)}]{vasmer2022morphing}%
  \BibitemOpen
  \bibfield  {author} {\bibinfo {author} {\bibfnamefont {M.}~\bibnamefont {Vasmer}}\ and\ \bibinfo {author} {\bibfnamefont {A.}~\bibnamefont {Kubica}},\ }\bibfield  {title} {\bibinfo {title} {Morphing quantum codes},\ }\href {https://doi.org/10.1103/PRXQuantum.3.030319} {\bibfield  {journal} {\bibinfo  {journal} {Phys. Rev. X Quantum}\ }\textbf {\bibinfo {volume} {3}},\ \bibinfo {pages} {030319} (\bibinfo {year} {2022})}\BibitemShut {NoStop}%
\bibitem [{\citenamefont {Shor}(1994)}]{shor1994algorithms}%
  \BibitemOpen
  \bibfield  {author} {\bibinfo {author} {\bibfnamefont {P.~W.}\ \bibnamefont {Shor}},\ }\bibfield  {title} {\bibinfo {title} {Algorithms for quantum computation: discrete logarithms and factoring},\ }in\ \href {https://doi.org/10.1109/SFCS.1994.365700} {\emph {\bibinfo {booktitle} {Proceedings 35th annual symposium on foundations of computer science}}}\ (\bibinfo {organization} {Ieee},\ \bibinfo {year} {1994})\ p.\ \bibinfo {pages} {124}\BibitemShut {NoStop}%
\bibitem [{\citenamefont {Grover}(1997)}]{grover1997quantum}%
  \BibitemOpen
  \bibfield  {author} {\bibinfo {author} {\bibfnamefont {L.~K.}\ \bibnamefont {Grover}},\ }\bibfield  {title} {\bibinfo {title} {Quantum computers can search arbitrarily large databases by a single query},\ }\href {https://doi.org/10.1103/PhysRevLett.79.4709} {\bibfield  {journal} {\bibinfo  {journal} {Phys. Rev. Lett.}\ }\textbf {\bibinfo {volume} {79}},\ \bibinfo {pages} {4709} (\bibinfo {year} {1997})}\BibitemShut {NoStop}%
\bibitem [{\citenamefont {Preskill}(2018)}]{preskill2018quantum}%
  \BibitemOpen
  \bibfield  {author} {\bibinfo {author} {\bibfnamefont {J.}~\bibnamefont {Preskill}},\ }\bibfield  {title} {\bibinfo {title} {Quantum computing in the {NISQ} era and beyond},\ }\href {https://doi.org/10.22331/q-2018-08-06-79} {\bibfield  {journal} {\bibinfo  {journal} {Quantum}\ }\textbf {\bibinfo {volume} {2}},\ \bibinfo {pages} {79} (\bibinfo {year} {2018})}\BibitemShut {NoStop}%
\bibitem [{\citenamefont {Gottesman}(1997)}]{gottesman1997stabilizer}%
  \BibitemOpen
  \bibfield  {author} {\bibinfo {author} {\bibfnamefont {D.}~\bibnamefont {Gottesman}},\ }\emph {\bibinfo {title} {Stabilizer Codes and Quantum Error Correction}},\ \href {https://thesis.library.caltech.edu/2900/2/THESIS.pdf} {Ph.D. thesis},\ \bibinfo  {school} {California Institute of Technology} (\bibinfo {year} {1997})\BibitemShut {NoStop}%
\bibitem [{\citenamefont {Nielsen}\ and\ \citenamefont {Chuang}(2010)}]{Nielsen_and_Chuang}%
  \BibitemOpen
  \bibfield  {author} {\bibinfo {author} {\bibfnamefont {M.~A.}\ \bibnamefont {Nielsen}}\ and\ \bibinfo {author} {\bibfnamefont {I.~L.}\ \bibnamefont {Chuang}},\ }\href {https://doi.org/10.1017/CBO9780511976667} {\emph {\bibinfo {title} {Quantum Computation and Quantum Information: 10th Anniversary Edition}}}\ (\bibinfo  {publisher} {Cambridge University Press},\ \bibinfo {year} {2010})\BibitemShut {NoStop}%
\bibitem [{\citenamefont {Aliferis}\ \emph {et~al.}(2006)\citenamefont {Aliferis}, \citenamefont {Gottesman},\ and\ \citenamefont {Preskill}}]{aliferis2005quantum}%
  \BibitemOpen
  \bibfield  {author} {\bibinfo {author} {\bibfnamefont {P.}~\bibnamefont {Aliferis}}, \bibinfo {author} {\bibfnamefont {D.}~\bibnamefont {Gottesman}},\ and\ \bibinfo {author} {\bibfnamefont {J.}~\bibnamefont {Preskill}},\ }\bibfield  {title} {\bibinfo {title} {Quantum accuracy threshold for concatenated distance-3 codes},\ }\href {https://www.rintonpress.com/journals/doi/QIC6.2-1.html} {\bibfield  {journal} {\bibinfo  {journal} {Quantum Inf. Comput.}\ }\textbf {\bibinfo {volume} {6}},\ \bibinfo {pages} {97} (\bibinfo {year} {2006})}\BibitemShut {NoStop}%
\bibitem [{\citenamefont {Knill}\ \emph {et~al.}(1998)\citenamefont {Knill}, \citenamefont {Laflamme},\ and\ \citenamefont {Zurek}}]{knill1998resilient}%
  \BibitemOpen
  \bibfield  {author} {\bibinfo {author} {\bibfnamefont {E.}~\bibnamefont {Knill}}, \bibinfo {author} {\bibfnamefont {R.}~\bibnamefont {Laflamme}},\ and\ \bibinfo {author} {\bibfnamefont {W.~H.}\ \bibnamefont {Zurek}},\ }\bibfield  {title} {\bibinfo {title} {Resilient quantum computation},\ }\href {https://doi.org/10.1126/science.279.5349.342} {\bibfield  {journal} {\bibinfo  {journal} {Science}\ }\textbf {\bibinfo {volume} {279}},\ \bibinfo {pages} {342} (\bibinfo {year} {1998})}\BibitemShut {NoStop}%
\bibitem [{\citenamefont {Aharonov}\ and\ \citenamefont {Ben-Or}(2008)}]{aharonov1997fault}%
  \BibitemOpen
  \bibfield  {author} {\bibinfo {author} {\bibfnamefont {D.}~\bibnamefont {Aharonov}}\ and\ \bibinfo {author} {\bibfnamefont {M.}~\bibnamefont {Ben-Or}},\ }\bibfield  {title} {\bibinfo {title} {Fault-tolerant quantum computation with constant error rate},\ }\href {https://doi.org/10.1137/S0097539799359385} {\bibfield  {journal} {\bibinfo  {journal} {SIAM Journal on Computing}\ }\textbf {\bibinfo {volume} {38}},\ \bibinfo {pages} {1207} (\bibinfo {year} {2008})}\BibitemShut {NoStop}%
\bibitem [{\citenamefont {Goto}(2016)}]{goto2016minimizing}%
  \BibitemOpen
  \bibfield  {author} {\bibinfo {author} {\bibfnamefont {H.}~\bibnamefont {Goto}},\ }\bibfield  {title} {\bibinfo {title} {Minimizing resource overheads for fault-tolerant preparation of encoded states of the {S}teane code},\ }\href {https://doi.org/10.1038/srep19578} {\bibfield  {journal} {\bibinfo  {journal} {Sci. Rep.}\ }\textbf {\bibinfo {volume} {6}},\ \bibinfo {pages} {1} (\bibinfo {year} {2016})}\BibitemShut {NoStop}%
\bibitem [{\citenamefont {Chamberland}\ and\ \citenamefont {Cross}(2019)}]{chamberland2019fault}%
  \BibitemOpen
  \bibfield  {author} {\bibinfo {author} {\bibfnamefont {C.}~\bibnamefont {Chamberland}}\ and\ \bibinfo {author} {\bibfnamefont {A.~W.}\ \bibnamefont {Cross}},\ }\bibfield  {title} {\bibinfo {title} {Fault-tolerant magic state preparation with flag qubits},\ }\href {https://doi.org/10.22331/q-2019-05-20-143} {\bibfield  {journal} {\bibinfo  {journal} {Quantum}\ }\textbf {\bibinfo {volume} {3}},\ \bibinfo {pages} {143} (\bibinfo {year} {2019})}\BibitemShut {NoStop}%
\bibitem [{\citenamefont {Bravyi}\ and\ \citenamefont {Kitaev}(2005)}]{bravyi2005universal}%
  \BibitemOpen
  \bibfield  {author} {\bibinfo {author} {\bibfnamefont {S.}~\bibnamefont {Bravyi}}\ and\ \bibinfo {author} {\bibfnamefont {A.}~\bibnamefont {Kitaev}},\ }\bibfield  {title} {\bibinfo {title} {Universal quantum computation with ideal {C}lifford gates and noisy ancillas},\ }\href {https://doi.org/10.1103/PhysRevA.71.022316} {\bibfield  {journal} {\bibinfo  {journal} {Phys. Rev. A}\ }\textbf {\bibinfo {volume} {71}},\ \bibinfo {pages} {022316} (\bibinfo {year} {2005})}\BibitemShut {NoStop}%
\bibitem [{\citenamefont {Gupta}\ \emph {et~al.}(2024)\citenamefont {Gupta}, \citenamefont {Sundaresan}, \citenamefont {Alexander}, \citenamefont {Wood}, \citenamefont {Merkel}, \citenamefont {Healy}, \citenamefont {Hillenbrand}, \citenamefont {Jochym-O’Connor}, \citenamefont {Wootton}, \citenamefont {Yoder} \emph {et~al.}}]{gupta2024encoding}%
  \BibitemOpen
  \bibfield  {author} {\bibinfo {author} {\bibfnamefont {R.~S.}\ \bibnamefont {Gupta}}, \bibinfo {author} {\bibfnamefont {N.}~\bibnamefont {Sundaresan}}, \bibinfo {author} {\bibfnamefont {T.}~\bibnamefont {Alexander}}, \bibinfo {author} {\bibfnamefont {C.~J.}\ \bibnamefont {Wood}}, \bibinfo {author} {\bibfnamefont {S.~T.}\ \bibnamefont {Merkel}}, \bibinfo {author} {\bibfnamefont {M.~B.}\ \bibnamefont {Healy}}, \bibinfo {author} {\bibfnamefont {M.}~\bibnamefont {Hillenbrand}}, \bibinfo {author} {\bibfnamefont {T.}~\bibnamefont {Jochym-O’Connor}}, \bibinfo {author} {\bibfnamefont {J.~R.}\ \bibnamefont {Wootton}}, \bibinfo {author} {\bibfnamefont {T.~J.}\ \bibnamefont {Yoder}}, \emph {et~al.},\ }\bibfield  {title} {\bibinfo {title} {Encoding a magic state with beyond break-even fidelity},\ }\href {https://doi.org/10.1038/s41586-023-06846-3} {\bibfield  {journal} {\bibinfo  {journal} {Nature}\ }\textbf {\bibinfo {volume} {625}},\ \bibinfo {pages} {259} (\bibinfo {year} {2024})}\BibitemShut {NoStop}%
\bibitem [{\citenamefont {Egan}\ \emph {et~al.}(2021)\citenamefont {Egan}, \citenamefont {Debroy}, \citenamefont {Noel}, \citenamefont {Risinger}, \citenamefont {Zhu}, \citenamefont {Biswas}, \citenamefont {Newman}, \citenamefont {Li}, \citenamefont {Brown}, \citenamefont {Cetina} \emph {et~al.}}]{egan2021fault}%
  \BibitemOpen
  \bibfield  {author} {\bibinfo {author} {\bibfnamefont {L.}~\bibnamefont {Egan}}, \bibinfo {author} {\bibfnamefont {D.~M.}\ \bibnamefont {Debroy}}, \bibinfo {author} {\bibfnamefont {C.}~\bibnamefont {Noel}}, \bibinfo {author} {\bibfnamefont {A.}~\bibnamefont {Risinger}}, \bibinfo {author} {\bibfnamefont {D.}~\bibnamefont {Zhu}}, \bibinfo {author} {\bibfnamefont {D.}~\bibnamefont {Biswas}}, \bibinfo {author} {\bibfnamefont {M.}~\bibnamefont {Newman}}, \bibinfo {author} {\bibfnamefont {M.}~\bibnamefont {Li}}, \bibinfo {author} {\bibfnamefont {K.~R.}\ \bibnamefont {Brown}}, \bibinfo {author} {\bibfnamefont {M.}~\bibnamefont {Cetina}}, \emph {et~al.},\ }\bibfield  {title} {\bibinfo {title} {Fault-tolerant control of an error-corrected qubit},\ }\href {https://doi.org/10.1038/s41586-021-03928-y} {\bibfield  {journal} {\bibinfo  {journal} {Nature}\ }\textbf {\bibinfo {volume} {598}},\ \bibinfo {pages} {281} (\bibinfo {year} {2021})}\BibitemShut {NoStop}%
\bibitem [{\citenamefont {Postler}\ \emph {et~al.}(2022)\citenamefont {Postler}, \citenamefont {Heu{\ss}en}, \citenamefont {Pogorelov}, \citenamefont {Rispler}, \citenamefont {Feldker}, \citenamefont {Meth}, \citenamefont {Marciniak}, \citenamefont {Stricker}, \citenamefont {Ringbauer}, \citenamefont {Blatt}, \citenamefont {Schindler}, \citenamefont {Müller},\ and\ \citenamefont {Monz}}]{postler2022demonstration}%
  \BibitemOpen
  \bibfield  {author} {\bibinfo {author} {\bibfnamefont {L.}~\bibnamefont {Postler}}, \bibinfo {author} {\bibfnamefont {S.}~\bibnamefont {Heu{\ss}en}}, \bibinfo {author} {\bibfnamefont {I.}~\bibnamefont {Pogorelov}}, \bibinfo {author} {\bibfnamefont {M.}~\bibnamefont {Rispler}}, \bibinfo {author} {\bibfnamefont {T.}~\bibnamefont {Feldker}}, \bibinfo {author} {\bibfnamefont {M.}~\bibnamefont {Meth}}, \bibinfo {author} {\bibfnamefont {C.~D.}\ \bibnamefont {Marciniak}}, \bibinfo {author} {\bibfnamefont {R.}~\bibnamefont {Stricker}}, \bibinfo {author} {\bibfnamefont {M.}~\bibnamefont {Ringbauer}}, \bibinfo {author} {\bibfnamefont {R.}~\bibnamefont {Blatt}}, \bibinfo {author} {\bibfnamefont {P.}~\bibnamefont {Schindler}}, \bibinfo {author} {\bibfnamefont {M.}~\bibnamefont {Müller}},\ and\ \bibinfo {author} {\bibfnamefont {T.}~\bibnamefont {Monz}},\ }\bibfield  {title} {\bibinfo {title} {{Demonstration of fault-tolerant universal quantum gate operations}},\ }\href {https://doi.org/10.1038/s41586-022-04721-1}
  {\bibfield  {journal} {\bibinfo  {journal} {Nature}\ }\textbf {\bibinfo {volume} {605}},\ \bibinfo {pages} {675} (\bibinfo {year} {2022})}\BibitemShut {NoStop}%
\bibitem [{\citenamefont {Gidney}\ and\ \citenamefont {Fowler}(2019)}]{gidney2019efficient}%
  \BibitemOpen
  \bibfield  {author} {\bibinfo {author} {\bibfnamefont {C.}~\bibnamefont {Gidney}}\ and\ \bibinfo {author} {\bibfnamefont {A.~G.}\ \bibnamefont {Fowler}},\ }\bibfield  {title} {\bibinfo {title} {Efficient magic state factories with a catalyzed $|{CCZ}\rangle$ to 2 $|{T}\rangle$ transformation},\ }\href {https://doi.org/10.22331/q-2019-04-30-135} {\bibfield  {journal} {\bibinfo  {journal} {Quantum}\ }\textbf {\bibinfo {volume} {3}},\ \bibinfo {pages} {135} (\bibinfo {year} {2019})}\BibitemShut {NoStop}%
\bibitem [{\citenamefont {Chao}\ and\ \citenamefont {Reichardt}(2018)}]{chao2018quantum}%
  \BibitemOpen
  \bibfield  {author} {\bibinfo {author} {\bibfnamefont {R.}~\bibnamefont {Chao}}\ and\ \bibinfo {author} {\bibfnamefont {B.~W.}\ \bibnamefont {Reichardt}},\ }\bibfield  {title} {\bibinfo {title} {Quantum error correction with only two extra qubits},\ }\href {https://doi.org/10.1103/PhysRevLett.121.050502} {\bibfield  {journal} {\bibinfo  {journal} {Phys. Rev. Lett.}\ }\textbf {\bibinfo {volume} {121}},\ \bibinfo {pages} {050502} (\bibinfo {year} {2018})}\BibitemShut {NoStop}%
\bibitem [{\citenamefont {Chamberland}\ and\ \citenamefont {Beverland}(2018)}]{chamberland2018flag}%
  \BibitemOpen
  \bibfield  {author} {\bibinfo {author} {\bibfnamefont {C.}~\bibnamefont {Chamberland}}\ and\ \bibinfo {author} {\bibfnamefont {M.~E.}\ \bibnamefont {Beverland}},\ }\bibfield  {title} {\bibinfo {title} {Flag fault-tolerant error correction with arbitrary distance codes},\ }\href {https://doi.org/10.22331/q-2018-02-08-53} {\bibfield  {journal} {\bibinfo  {journal} {Quantum}\ }\textbf {\bibinfo {volume} {2}},\ \bibinfo {pages} {53} (\bibinfo {year} {2018})}\BibitemShut {NoStop}%
\bibitem [{\citenamefont {Butt}\ \emph {et~al.}(2023)\citenamefont {Butt}, \citenamefont {Heu{\ss}en}, \citenamefont {Rispler},\ and\ \citenamefont {M{\"u}ller}}]{butt2023fault}%
  \BibitemOpen
  \bibfield  {author} {\bibinfo {author} {\bibfnamefont {F.}~\bibnamefont {Butt}}, \bibinfo {author} {\bibfnamefont {S.}~\bibnamefont {Heu{\ss}en}}, \bibinfo {author} {\bibfnamefont {M.}~\bibnamefont {Rispler}},\ and\ \bibinfo {author} {\bibfnamefont {M.}~\bibnamefont {M{\"u}ller}},\ }\bibfield  {title} {\bibinfo {title} {Fault-tolerant code switching protocols for near-term quantum processors},\ }\href {https://doi.org/10.48550/arXiv.2306.17686} {\bibfield  {journal} {\bibinfo  {journal} {arXiv preprint arXiv:2306.17686}\ } (\bibinfo {year} {2023})}\BibitemShut {NoStop}%
\bibitem [{\citenamefont {Bombin}\ and\ \citenamefont {Martin-Delgado}(2006)}]{bombin2006topological}%
  \BibitemOpen
  \bibfield  {author} {\bibinfo {author} {\bibfnamefont {H.}~\bibnamefont {Bombin}}\ and\ \bibinfo {author} {\bibfnamefont {M.~A.}\ \bibnamefont {Martin-Delgado}},\ }\bibfield  {title} {\bibinfo {title} {Topological quantum distillation},\ }\href {https://doi.org/10.1103/PhysRevLett.97.180501} {\bibfield  {journal} {\bibinfo  {journal} {Phys. Rev. Lett.}\ }\textbf {\bibinfo {volume} {97}},\ \bibinfo {pages} {180501} (\bibinfo {year} {2006})}\BibitemShut {NoStop}%
\bibitem [{\citenamefont {Poulin}(2005)}]{poulin2005stabilizer}%
  \BibitemOpen
  \bibfield  {author} {\bibinfo {author} {\bibfnamefont {D.}~\bibnamefont {Poulin}},\ }\bibfield  {title} {\bibinfo {title} {Stabilizer formalism for operator quantum error correction},\ }\href {https://doi.org/10.1103/PhysRevLett.95.230504} {\bibfield  {journal} {\bibinfo  {journal} {Phys. Rev. Lett.}\ }\textbf {\bibinfo {volume} {95}},\ \bibinfo {pages} {230504} (\bibinfo {year} {2005})}\BibitemShut {NoStop}%
\bibitem [{\citenamefont {Hilder}\ \emph {et~al.}(2022)\citenamefont {Hilder}, \citenamefont {Pijn}, \citenamefont {Onishchenko}, \citenamefont {Stahl}, \citenamefont {Orth}, \citenamefont {Lekitsch}, \citenamefont {Rodriguez-Blanco}, \citenamefont {M{\"u}ller}, \citenamefont {Schmidt-Kaler},\ and\ \citenamefont {Poschinger}}]{hilder2022fault}%
  \BibitemOpen
  \bibfield  {author} {\bibinfo {author} {\bibfnamefont {J.}~\bibnamefont {Hilder}}, \bibinfo {author} {\bibfnamefont {D.}~\bibnamefont {Pijn}}, \bibinfo {author} {\bibfnamefont {O.}~\bibnamefont {Onishchenko}}, \bibinfo {author} {\bibfnamefont {A.}~\bibnamefont {Stahl}}, \bibinfo {author} {\bibfnamefont {M.}~\bibnamefont {Orth}}, \bibinfo {author} {\bibfnamefont {B.}~\bibnamefont {Lekitsch}}, \bibinfo {author} {\bibfnamefont {A.}~\bibnamefont {Rodriguez-Blanco}}, \bibinfo {author} {\bibfnamefont {M.}~\bibnamefont {M{\"u}ller}}, \bibinfo {author} {\bibfnamefont {F.}~\bibnamefont {Schmidt-Kaler}},\ and\ \bibinfo {author} {\bibfnamefont {U.}~\bibnamefont {Poschinger}},\ }\bibfield  {title} {\bibinfo {title} {Fault-tolerant parity readout on a shuttling-based trapped-ion quantum computer},\ }\href {https://doi.org/10.1103/PhysRevX.12.011032} {\bibfield  {journal} {\bibinfo  {journal} {Phys. Rev. X}\ }\textbf {\bibinfo {volume} {12}},\ \bibinfo {pages} {011032} (\bibinfo {year} {2022})}\BibitemShut {NoStop}%
\bibitem [{\citenamefont {S{\o}rensen}\ and\ \citenamefont {M{\o}lmer}(2000)}]{sorensen2000entanglement}%
  \BibitemOpen
  \bibfield  {author} {\bibinfo {author} {\bibfnamefont {A.}~\bibnamefont {S{\o}rensen}}\ and\ \bibinfo {author} {\bibfnamefont {K.}~\bibnamefont {M{\o}lmer}},\ }\bibfield  {title} {\bibinfo {title} {Entanglement and quantum computation with ions in thermal motion},\ }\href {https://doi.org/10.1103/PhysRevA.62.022311} {\bibfield  {journal} {\bibinfo  {journal} {Phys. Rev. A}\ }\textbf {\bibinfo {volume} {62}},\ \bibinfo {pages} {022311} (\bibinfo {year} {2000})}\BibitemShut {NoStop}%
\bibitem [{\citenamefont {Pogorelov}\ \emph {et~al.}(2021)\citenamefont {Pogorelov}, \citenamefont {Feldker}, \citenamefont {Marciniak}, \citenamefont {Postler}, \citenamefont {Jacob}, \citenamefont {Krieglsteiner}, \citenamefont {Podlesnic}, \citenamefont {Meth}, \citenamefont {Negnevitsky}, \citenamefont {Stadler}, \citenamefont {H\"ofer}, \citenamefont {W\"achter}, \citenamefont {Lakhmanskiy}, \citenamefont {Blatt}, \citenamefont {Schindler},\ and\ \citenamefont {Monz}}]{pogorelov2021compact}%
  \BibitemOpen
  \bibfield  {author} {\bibinfo {author} {\bibfnamefont {I.}~\bibnamefont {Pogorelov}}, \bibinfo {author} {\bibfnamefont {T.}~\bibnamefont {Feldker}}, \bibinfo {author} {\bibfnamefont {C.~D.}\ \bibnamefont {Marciniak}}, \bibinfo {author} {\bibfnamefont {L.}~\bibnamefont {Postler}}, \bibinfo {author} {\bibfnamefont {G.}~\bibnamefont {Jacob}}, \bibinfo {author} {\bibfnamefont {O.}~\bibnamefont {Krieglsteiner}}, \bibinfo {author} {\bibfnamefont {V.}~\bibnamefont {Podlesnic}}, \bibinfo {author} {\bibfnamefont {M.}~\bibnamefont {Meth}}, \bibinfo {author} {\bibfnamefont {V.}~\bibnamefont {Negnevitsky}}, \bibinfo {author} {\bibfnamefont {M.}~\bibnamefont {Stadler}}, \bibinfo {author} {\bibfnamefont {B.}~\bibnamefont {H\"ofer}}, \bibinfo {author} {\bibfnamefont {C.}~\bibnamefont {W\"achter}}, \bibinfo {author} {\bibfnamefont {K.}~\bibnamefont {Lakhmanskiy}}, \bibinfo {author} {\bibfnamefont {R.}~\bibnamefont {Blatt}}, \bibinfo {author} {\bibfnamefont {P.}~\bibnamefont {Schindler}},\ and\ \bibinfo {author}
  {\bibfnamefont {T.}~\bibnamefont {Monz}},\ }\bibfield  {title} {\bibinfo {title} {Compact ion-trap quantum computing demonstrator},\ }\href {https://doi.org/10.1103/PRXQuantum.2.020343} {\bibfield  {journal} {\bibinfo  {journal} {Phys. Rev. X Quantum}\ }\textbf {\bibinfo {volume} {2}},\ \bibinfo {pages} {020343} (\bibinfo {year} {2021})}\BibitemShut {NoStop}%
\bibitem [{\citenamefont {Heu{\ss}en}\ \emph {et~al.}(2023)\citenamefont {Heu{\ss}en}, \citenamefont {Postler}, \citenamefont {Rispler}, \citenamefont {Pogorelov}, \citenamefont {Marciniak}, \citenamefont {Monz}, \citenamefont {Schindler},\ and\ \citenamefont {M{\"u}ller}}]{heussen2023strategies}%
  \BibitemOpen
  \bibfield  {author} {\bibinfo {author} {\bibfnamefont {S.}~\bibnamefont {Heu{\ss}en}}, \bibinfo {author} {\bibfnamefont {L.}~\bibnamefont {Postler}}, \bibinfo {author} {\bibfnamefont {M.}~\bibnamefont {Rispler}}, \bibinfo {author} {\bibfnamefont {I.}~\bibnamefont {Pogorelov}}, \bibinfo {author} {\bibfnamefont {C.~D.}\ \bibnamefont {Marciniak}}, \bibinfo {author} {\bibfnamefont {T.}~\bibnamefont {Monz}}, \bibinfo {author} {\bibfnamefont {P.}~\bibnamefont {Schindler}},\ and\ \bibinfo {author} {\bibfnamefont {M.}~\bibnamefont {M{\"u}ller}},\ }\bibfield  {title} {\bibinfo {title} {{Strategies for a practical advantage of fault-tolerant circuit design in noisy trapped-ion quantum computers}},\ }\href {https://doi.org/10.1103/PhysRevA.107.042422} {\bibfield  {journal} {\bibinfo  {journal} {Phys. Rev. A}\ }\textbf {\bibinfo {volume} {107}},\ \bibinfo {pages} {042422} (\bibinfo {year} {2023})}\BibitemShut {NoStop}%
\bibitem [{\citenamefont {Postler}\ \emph {et~al.}(2023)\citenamefont {Postler}, \citenamefont {Butt}, \citenamefont {Pogorelov}, \citenamefont {Marciniak}, \citenamefont {Heu{\ss}en}, \citenamefont {Blatt}, \citenamefont {Schindler}, \citenamefont {Rispler}, \citenamefont {M{\"u}ller},\ and\ \citenamefont {Monz}}]{postler2023demonstration}%
  \BibitemOpen
  \bibfield  {author} {\bibinfo {author} {\bibfnamefont {L.}~\bibnamefont {Postler}}, \bibinfo {author} {\bibfnamefont {F.}~\bibnamefont {Butt}}, \bibinfo {author} {\bibfnamefont {I.}~\bibnamefont {Pogorelov}}, \bibinfo {author} {\bibfnamefont {C.~D.}\ \bibnamefont {Marciniak}}, \bibinfo {author} {\bibfnamefont {S.}~\bibnamefont {Heu{\ss}en}}, \bibinfo {author} {\bibfnamefont {R.}~\bibnamefont {Blatt}}, \bibinfo {author} {\bibfnamefont {P.}~\bibnamefont {Schindler}}, \bibinfo {author} {\bibfnamefont {M.}~\bibnamefont {Rispler}}, \bibinfo {author} {\bibfnamefont {M.}~\bibnamefont {M{\"u}ller}},\ and\ \bibinfo {author} {\bibfnamefont {T.}~\bibnamefont {Monz}},\ }\bibfield  {title} {\bibinfo {title} {Demonstration of fault-tolerant {S}teane quantum error correction},\ }\href {https://doi.org/10.48550/arXiv.2312.09745} {\bibfield  {journal} {\bibinfo  {journal} {arXiv preprint arXiv:2312.09745}\ } (\bibinfo {year} {2023})}\BibitemShut {NoStop}%
\bibitem [{\citenamefont {Harty}\ \emph {et~al.}(2014)\citenamefont {Harty}, \citenamefont {Allcock}, \citenamefont {Ballance}, \citenamefont {Guidoni}, \citenamefont {Janacek}, \citenamefont {Linke}, \citenamefont {Stacey},\ and\ \citenamefont {Lucas}}]{harty2014high}%
  \BibitemOpen
  \bibfield  {author} {\bibinfo {author} {\bibfnamefont {T.~P.}\ \bibnamefont {Harty}}, \bibinfo {author} {\bibfnamefont {D.~T.~C.}\ \bibnamefont {Allcock}}, \bibinfo {author} {\bibfnamefont {C.~J.}\ \bibnamefont {Ballance}}, \bibinfo {author} {\bibfnamefont {L.}~\bibnamefont {Guidoni}}, \bibinfo {author} {\bibfnamefont {H.~A.}\ \bibnamefont {Janacek}}, \bibinfo {author} {\bibfnamefont {N.~M.}\ \bibnamefont {Linke}}, \bibinfo {author} {\bibfnamefont {D.~N.}\ \bibnamefont {Stacey}},\ and\ \bibinfo {author} {\bibfnamefont {D.~M.}\ \bibnamefont {Lucas}},\ }\bibfield  {title} {\bibinfo {title} {{High-fidelity preparation, gates, memory, and readout of a trapped-ion quantum bit}},\ }\href {https://doi.org/10.1103/PhysRevLett.113.220501} {\bibfield  {journal} {\bibinfo  {journal} {Phys. Rev. Lett.}\ }\textbf {\bibinfo {volume} {113}},\ \bibinfo {pages} {220501} (\bibinfo {year} {2014})}\BibitemShut {NoStop}%
\bibitem [{\citenamefont {Ruster}\ \emph {et~al.}(2016)\citenamefont {Ruster}, \citenamefont {Schmiegelow}, \citenamefont {Kaufmann}, \citenamefont {Warschburger}, \citenamefont {Schmidt-Kaler},\ and\ \citenamefont {Poschinger}}]{ruster2016long}%
  \BibitemOpen
  \bibfield  {author} {\bibinfo {author} {\bibfnamefont {T.}~\bibnamefont {Ruster}}, \bibinfo {author} {\bibfnamefont {C.~T.}\ \bibnamefont {Schmiegelow}}, \bibinfo {author} {\bibfnamefont {H.}~\bibnamefont {Kaufmann}}, \bibinfo {author} {\bibfnamefont {C.}~\bibnamefont {Warschburger}}, \bibinfo {author} {\bibfnamefont {F.}~\bibnamefont {Schmidt-Kaler}},\ and\ \bibinfo {author} {\bibfnamefont {U.~G.}\ \bibnamefont {Poschinger}},\ }\bibfield  {title} {\bibinfo {title} {{A long-lived Zeeman trapped-ion qubit}},\ }\href {https://doi.org/10.1007/s00340-016-6527-4} {\bibfield  {journal} {\bibinfo  {journal} {Appl. Phys. B}\ }\textbf {\bibinfo {volume} {122}},\ \bibinfo {pages} {254} (\bibinfo {year} {2016})}\BibitemShut {NoStop}%
\bibitem [{\citenamefont {Wang}\ \emph {et~al.}(2021)\citenamefont {Wang}, \citenamefont {Luan}, \citenamefont {Qiao}, \citenamefont {Um}, \citenamefont {Zhang}, \citenamefont {Wang}, \citenamefont {Yuan}, \citenamefont {Gu}, \citenamefont {Zhang},\ and\ \citenamefont {Kim}}]{wang2021single}%
  \BibitemOpen
  \bibfield  {author} {\bibinfo {author} {\bibfnamefont {P.}~\bibnamefont {Wang}}, \bibinfo {author} {\bibfnamefont {C.-Y.}\ \bibnamefont {Luan}}, \bibinfo {author} {\bibfnamefont {M.}~\bibnamefont {Qiao}}, \bibinfo {author} {\bibfnamefont {M.}~\bibnamefont {Um}}, \bibinfo {author} {\bibfnamefont {J.}~\bibnamefont {Zhang}}, \bibinfo {author} {\bibfnamefont {Y.}~\bibnamefont {Wang}}, \bibinfo {author} {\bibfnamefont {X.}~\bibnamefont {Yuan}}, \bibinfo {author} {\bibfnamefont {M.}~\bibnamefont {Gu}}, \bibinfo {author} {\bibfnamefont {J.}~\bibnamefont {Zhang}},\ and\ \bibinfo {author} {\bibfnamefont {K.}~\bibnamefont {Kim}},\ }\bibfield  {title} {\bibinfo {title} {{Single ion qubit with estimated coherence time exceeding one hour}},\ }\href {https://doi.org/10.1038/s41467-020-20330-w} {\bibfield  {journal} {\bibinfo  {journal} {Nat. Commun.}\ }\textbf {\bibinfo {volume} {12}},\ \bibinfo {pages} {233} (\bibinfo {year} {2021})}\BibitemShut {NoStop}%
\bibitem [{\citenamefont {Wimperis}(1989)}]{wimperis1989composite}%
  \BibitemOpen
  \bibfield  {author} {\bibinfo {author} {\bibfnamefont {S.}~\bibnamefont {Wimperis}},\ }\bibfield  {title} {\bibinfo {title} {{Composite pulses with rectangular excitation and inversion profiles}},\ }\href {https://doi.org/https://doi.org/10.1016/0022-2364(89)90346-6} {\bibfield  {journal} {\bibinfo  {journal} {J. Magn. Reson.}\ }\textbf {\bibinfo {volume} {83}},\ \bibinfo {pages} {509} (\bibinfo {year} {1989})}\BibitemShut {NoStop}%
\bibitem [{\citenamefont {Wimperis}(1994)}]{wimperis1994broadband}%
  \BibitemOpen
  \bibfield  {author} {\bibinfo {author} {\bibfnamefont {S.}~\bibnamefont {Wimperis}},\ }\bibfield  {title} {\bibinfo {title} {{Broadband, narrowband, and passband composite pulses for use in advanced NMR experiments}},\ }\href {https://doi.org/https://doi.org/10.1006/jmra.1994.1159} {\bibfield  {journal} {\bibinfo  {journal} {J. Magn. Reson.}\ }\textbf {\bibinfo {volume} {109}},\ \bibinfo {pages} {221} (\bibinfo {year} {1994})}\BibitemShut {NoStop}%
\bibitem [{\citenamefont {Bonilla~Ataides}\ \emph {et~al.}(2021)\citenamefont {Bonilla~Ataides}, \citenamefont {Tuckett}, \citenamefont {Bartlett}, \citenamefont {Flammia},\ and\ \citenamefont {Brown}}]{bonilla2021xzzx}%
  \BibitemOpen
  \bibfield  {author} {\bibinfo {author} {\bibfnamefont {J.~P.}\ \bibnamefont {Bonilla~Ataides}}, \bibinfo {author} {\bibfnamefont {D.~K.}\ \bibnamefont {Tuckett}}, \bibinfo {author} {\bibfnamefont {S.~D.}\ \bibnamefont {Bartlett}}, \bibinfo {author} {\bibfnamefont {S.~T.}\ \bibnamefont {Flammia}},\ and\ \bibinfo {author} {\bibfnamefont {B.~J.}\ \bibnamefont {Brown}},\ }\bibfield  {title} {\bibinfo {title} {The {XZZX} surface code},\ }\href {https://doi.org/10.1038/s41467-021-22274-1} {\bibfield  {journal} {\bibinfo  {journal} {Nat. Commun.}\ }\textbf {\bibinfo {volume} {12}},\ \bibinfo {pages} {2172} (\bibinfo {year} {2021})}\BibitemShut {NoStop}%
\bibitem [{\citenamefont {Xu}\ \emph {et~al.}(2023)\citenamefont {Xu}, \citenamefont {Mannucci}, \citenamefont {Seif}, \citenamefont {Kubica}, \citenamefont {Flammia},\ and\ \citenamefont {Jiang}}]{xu2023tailored}%
  \BibitemOpen
  \bibfield  {author} {\bibinfo {author} {\bibfnamefont {Q.}~\bibnamefont {Xu}}, \bibinfo {author} {\bibfnamefont {N.}~\bibnamefont {Mannucci}}, \bibinfo {author} {\bibfnamefont {A.}~\bibnamefont {Seif}}, \bibinfo {author} {\bibfnamefont {A.}~\bibnamefont {Kubica}}, \bibinfo {author} {\bibfnamefont {S.~T.}\ \bibnamefont {Flammia}},\ and\ \bibinfo {author} {\bibfnamefont {L.}~\bibnamefont {Jiang}},\ }\bibfield  {title} {\bibinfo {title} {Tailored {XZZX} codes for biased noise},\ }\href {https://doi.org/10.1103/PhysRevResearch.5.013035} {\bibfield  {journal} {\bibinfo  {journal} {Phys. Rev. Res.}\ }\textbf {\bibinfo {volume} {5}},\ \bibinfo {pages} {013035} (\bibinfo {year} {2023})}\BibitemShut {NoStop}%
\bibitem [{\citenamefont {Huang}\ \emph {et~al.}(2023{\natexlab{a}})\citenamefont {Huang}, \citenamefont {Pesah}, \citenamefont {Chubb}, \citenamefont {Vasmer},\ and\ \citenamefont {Dua}}]{huang2023tailoring}%
  \BibitemOpen
  \bibfield  {author} {\bibinfo {author} {\bibfnamefont {E.}~\bibnamefont {Huang}}, \bibinfo {author} {\bibfnamefont {A.}~\bibnamefont {Pesah}}, \bibinfo {author} {\bibfnamefont {C.~T.}\ \bibnamefont {Chubb}}, \bibinfo {author} {\bibfnamefont {M.}~\bibnamefont {Vasmer}},\ and\ \bibinfo {author} {\bibfnamefont {A.}~\bibnamefont {Dua}},\ }\bibfield  {title} {\bibinfo {title} {Tailoring three-dimensional topological codes for biased noise},\ }\href {https://doi.org/10.1103/PRXQuantum.4.030338} {\bibfield  {journal} {\bibinfo  {journal} {Phys. Rev. X Quantum}\ }\textbf {\bibinfo {volume} {4}},\ \bibinfo {pages} {030338} (\bibinfo {year} {2023}{\natexlab{a}})}\BibitemShut {NoStop}%
\bibitem [{\citenamefont {Pal}\ \emph {et~al.}(2022)\citenamefont {Pal}, \citenamefont {Schindler}, \citenamefont {Erhard}, \citenamefont {Rivas}, \citenamefont {Martin-Delgado}, \citenamefont {Blatt}, \citenamefont {Monz},\ and\ \citenamefont {M{\"u}ller}}]{pal2022relaxation}%
  \BibitemOpen
  \bibfield  {author} {\bibinfo {author} {\bibfnamefont {A.~K.}\ \bibnamefont {Pal}}, \bibinfo {author} {\bibfnamefont {P.}~\bibnamefont {Schindler}}, \bibinfo {author} {\bibfnamefont {A.}~\bibnamefont {Erhard}}, \bibinfo {author} {\bibfnamefont {{\'A}.}~\bibnamefont {Rivas}}, \bibinfo {author} {\bibfnamefont {M.-A.}\ \bibnamefont {Martin-Delgado}}, \bibinfo {author} {\bibfnamefont {R.}~\bibnamefont {Blatt}}, \bibinfo {author} {\bibfnamefont {T.}~\bibnamefont {Monz}},\ and\ \bibinfo {author} {\bibfnamefont {M.}~\bibnamefont {M{\"u}ller}},\ }\bibfield  {title} {\bibinfo {title} {Relaxation times do not capture logical qubit dynamics},\ }\href {https://doi.org/10.22331/q-2022-01-24-632} {\bibfield  {journal} {\bibinfo  {journal} {Quantum}\ }\textbf {\bibinfo {volume} {6}},\ \bibinfo {pages} {632} (\bibinfo {year} {2022})}\BibitemShut {NoStop}%
\bibitem [{\citenamefont {Krinner}\ \emph {et~al.}(2022)\citenamefont {Krinner}, \citenamefont {Lacroix}, \citenamefont {Remm}, \citenamefont {Di~Paolo}, \citenamefont {Genois}, \citenamefont {Leroux}, \citenamefont {Hellings}, \citenamefont {Lazar}, \citenamefont {Swiadek}, \citenamefont {Herrmann} \emph {et~al.}}]{krinner2022realizing}%
  \BibitemOpen
  \bibfield  {author} {\bibinfo {author} {\bibfnamefont {S.}~\bibnamefont {Krinner}}, \bibinfo {author} {\bibfnamefont {N.}~\bibnamefont {Lacroix}}, \bibinfo {author} {\bibfnamefont {A.}~\bibnamefont {Remm}}, \bibinfo {author} {\bibfnamefont {A.}~\bibnamefont {Di~Paolo}}, \bibinfo {author} {\bibfnamefont {E.}~\bibnamefont {Genois}}, \bibinfo {author} {\bibfnamefont {C.}~\bibnamefont {Leroux}}, \bibinfo {author} {\bibfnamefont {C.}~\bibnamefont {Hellings}}, \bibinfo {author} {\bibfnamefont {S.}~\bibnamefont {Lazar}}, \bibinfo {author} {\bibfnamefont {F.}~\bibnamefont {Swiadek}}, \bibinfo {author} {\bibfnamefont {J.}~\bibnamefont {Herrmann}}, \emph {et~al.},\ }\bibfield  {title} {\bibinfo {title} {Realizing repeated quantum error correction in a distance-three surface code},\ }\href {https://doi.org/10.1038/s41586-022-04566-8} {\bibfield  {journal} {\bibinfo  {journal} {Nature}\ }\textbf {\bibinfo {volume} {605}},\ \bibinfo {pages} {669} (\bibinfo {year} {2022})}\BibitemShut {NoStop}%
\bibitem [{\citenamefont {Kelly}\ \emph {et~al.}(2015)\citenamefont {Kelly}, \citenamefont {Barends}, \citenamefont {Fowler}, \citenamefont {Megrant}, \citenamefont {Jeffrey}, \citenamefont {White}, \citenamefont {Sank}, \citenamefont {Mutus}, \citenamefont {Campbell}, \citenamefont {Chen} \emph {et~al.}}]{kelly2015state}%
  \BibitemOpen
  \bibfield  {author} {\bibinfo {author} {\bibfnamefont {J.}~\bibnamefont {Kelly}}, \bibinfo {author} {\bibfnamefont {R.}~\bibnamefont {Barends}}, \bibinfo {author} {\bibfnamefont {A.~G.}\ \bibnamefont {Fowler}}, \bibinfo {author} {\bibfnamefont {A.}~\bibnamefont {Megrant}}, \bibinfo {author} {\bibfnamefont {E.}~\bibnamefont {Jeffrey}}, \bibinfo {author} {\bibfnamefont {T.~C.}\ \bibnamefont {White}}, \bibinfo {author} {\bibfnamefont {D.}~\bibnamefont {Sank}}, \bibinfo {author} {\bibfnamefont {J.~Y.}\ \bibnamefont {Mutus}}, \bibinfo {author} {\bibfnamefont {B.}~\bibnamefont {Campbell}}, \bibinfo {author} {\bibfnamefont {Y.}~\bibnamefont {Chen}}, \emph {et~al.},\ }\bibfield  {title} {\bibinfo {title} {State preservation by repetitive error detection in a superconducting quantum circuit},\ }\href {https://doi.org/10.1038/nature14270} {\bibfield  {journal} {\bibinfo  {journal} {Nature}\ }\textbf {\bibinfo {volume} {519}},\ \bibinfo {pages} {66} (\bibinfo {year} {2015})}\BibitemShut {NoStop}%
\bibitem [{\citenamefont {Zhao}\ \emph {et~al.}(2022)\citenamefont {Zhao}, \citenamefont {Ye}, \citenamefont {Huang}, \citenamefont {Zhang}, \citenamefont {Wu}, \citenamefont {Guan}, \citenamefont {Zhu}, \citenamefont {Wei}, \citenamefont {He}, \citenamefont {Cao} \emph {et~al.}}]{zhao2022realization}%
  \BibitemOpen
  \bibfield  {author} {\bibinfo {author} {\bibfnamefont {Y.}~\bibnamefont {Zhao}}, \bibinfo {author} {\bibfnamefont {Y.}~\bibnamefont {Ye}}, \bibinfo {author} {\bibfnamefont {H.-L.}\ \bibnamefont {Huang}}, \bibinfo {author} {\bibfnamefont {Y.}~\bibnamefont {Zhang}}, \bibinfo {author} {\bibfnamefont {D.}~\bibnamefont {Wu}}, \bibinfo {author} {\bibfnamefont {H.}~\bibnamefont {Guan}}, \bibinfo {author} {\bibfnamefont {Q.}~\bibnamefont {Zhu}}, \bibinfo {author} {\bibfnamefont {Z.}~\bibnamefont {Wei}}, \bibinfo {author} {\bibfnamefont {T.}~\bibnamefont {He}}, \bibinfo {author} {\bibfnamefont {S.}~\bibnamefont {Cao}}, \emph {et~al.},\ }\bibfield  {title} {\bibinfo {title} {Realization of an error-correcting surface code with superconducting qubits},\ }\href {https://doi.org/10.1103/PhysRevLett.129.030501} {\bibfield  {journal} {\bibinfo  {journal} {Phys. Rev. Lett.}\ }\textbf {\bibinfo {volume} {129}},\ \bibinfo {pages} {030501} (\bibinfo {year} {2022})}\BibitemShut {NoStop}%
\bibitem [{\citenamefont {AI}(2021)}]{google2021exponential}%
  \BibitemOpen
  \bibfield  {author} {\bibinfo {author} {\bibfnamefont {G.~Q.}\ \bibnamefont {AI}},\ }\bibfield  {title} {\bibinfo {title} {Exponential suppression of bit or phase errors with cyclic error correction},\ }\href {https://doi.org/10.1038/s41586-021-03588-y} {\bibfield  {journal} {\bibinfo  {journal} {Nature}\ }\textbf {\bibinfo {volume} {595}},\ \bibinfo {pages} {383} (\bibinfo {year} {2021})}\BibitemShut {NoStop}%
\bibitem [{\citenamefont {Satzinger}\ \emph {et~al.}(2021)\citenamefont {Satzinger}, \citenamefont {Liu}, \citenamefont {Smith}, \citenamefont {Knapp}, \citenamefont {Newman}, \citenamefont {Jones}, \citenamefont {Chen}, \citenamefont {Quintana}, \citenamefont {Mi}, \citenamefont {Dunsworth} \emph {et~al.}}]{satzinger2021realizing}%
  \BibitemOpen
  \bibfield  {author} {\bibinfo {author} {\bibfnamefont {K.}~\bibnamefont {Satzinger}}, \bibinfo {author} {\bibfnamefont {Y.-J.}\ \bibnamefont {Liu}}, \bibinfo {author} {\bibfnamefont {A.}~\bibnamefont {Smith}}, \bibinfo {author} {\bibfnamefont {C.}~\bibnamefont {Knapp}}, \bibinfo {author} {\bibfnamefont {M.}~\bibnamefont {Newman}}, \bibinfo {author} {\bibfnamefont {C.}~\bibnamefont {Jones}}, \bibinfo {author} {\bibfnamefont {Z.}~\bibnamefont {Chen}}, \bibinfo {author} {\bibfnamefont {C.}~\bibnamefont {Quintana}}, \bibinfo {author} {\bibfnamefont {X.}~\bibnamefont {Mi}}, \bibinfo {author} {\bibfnamefont {A.}~\bibnamefont {Dunsworth}}, \emph {et~al.},\ }\bibfield  {title} {\bibinfo {title} {Realizing topologically ordered states on a quantum processor},\ }\href {https://doi.org/10.1126/science.abi8378} {\bibfield  {journal} {\bibinfo  {journal} {Science}\ }\textbf {\bibinfo {volume} {374}},\ \bibinfo {pages} {1237} (\bibinfo {year} {2021})}\BibitemShut {NoStop}%
\bibitem [{\citenamefont {Takita}\ \emph {et~al.}(2017)\citenamefont {Takita}, \citenamefont {Cross}, \citenamefont {C{\'o}rcoles}, \citenamefont {Chow},\ and\ \citenamefont {Gambetta}}]{takita2017experimental}%
  \BibitemOpen
  \bibfield  {author} {\bibinfo {author} {\bibfnamefont {M.}~\bibnamefont {Takita}}, \bibinfo {author} {\bibfnamefont {A.~W.}\ \bibnamefont {Cross}}, \bibinfo {author} {\bibfnamefont {A.~D.}\ \bibnamefont {C{\'o}rcoles}}, \bibinfo {author} {\bibfnamefont {J.~M.}\ \bibnamefont {Chow}},\ and\ \bibinfo {author} {\bibfnamefont {J.~M.}\ \bibnamefont {Gambetta}},\ }\bibfield  {title} {\bibinfo {title} {Experimental demonstration of fault-tolerant state preparation with superconducting qubits},\ }\href {https://doi.org/10.1103/PhysRevLett.119.180501} {\bibfield  {journal} {\bibinfo  {journal} {Phys. Rev. Lett.}\ }\textbf {\bibinfo {volume} {119}},\ \bibinfo {pages} {180501} (\bibinfo {year} {2017})}\BibitemShut {NoStop}%
\bibitem [{\citenamefont {Pino}\ \emph {et~al.}(2021)\citenamefont {Pino}, \citenamefont {Dreiling}, \citenamefont {Figgatt}, \citenamefont {Gaebler}, \citenamefont {Moses}, \citenamefont {Allman}, \citenamefont {Baldwin}, \citenamefont {Foss-Feig}, \citenamefont {Hayes}, \citenamefont {Mayer} \emph {et~al.}}]{pino2021demonstration}%
  \BibitemOpen
  \bibfield  {author} {\bibinfo {author} {\bibfnamefont {J.~M.}\ \bibnamefont {Pino}}, \bibinfo {author} {\bibfnamefont {J.~M.}\ \bibnamefont {Dreiling}}, \bibinfo {author} {\bibfnamefont {C.}~\bibnamefont {Figgatt}}, \bibinfo {author} {\bibfnamefont {J.~P.}\ \bibnamefont {Gaebler}}, \bibinfo {author} {\bibfnamefont {S.~A.}\ \bibnamefont {Moses}}, \bibinfo {author} {\bibfnamefont {M.}~\bibnamefont {Allman}}, \bibinfo {author} {\bibfnamefont {C.}~\bibnamefont {Baldwin}}, \bibinfo {author} {\bibfnamefont {M.}~\bibnamefont {Foss-Feig}}, \bibinfo {author} {\bibfnamefont {D.}~\bibnamefont {Hayes}}, \bibinfo {author} {\bibfnamefont {K.}~\bibnamefont {Mayer}}, \emph {et~al.},\ }\bibfield  {title} {\bibinfo {title} {Demonstration of the trapped-ion quantum {CCD} computer architecture},\ }\href {https://doi.org/10.1038/s41586-021-03318-4} {\bibfield  {journal} {\bibinfo  {journal} {Nature}\ }\textbf {\bibinfo {volume} {592}},\ \bibinfo {pages} {209} (\bibinfo {year} {2021})}\BibitemShut {NoStop}%
\bibitem [{\citenamefont {Ryan-Anderson}\ \emph {et~al.}(2021)\citenamefont {Ryan-Anderson}, \citenamefont {Bohnet}, \citenamefont {Lee}, \citenamefont {Gresh}, \citenamefont {Hankin}, \citenamefont {Gaebler}, \citenamefont {Francois}, \citenamefont {Chernoguzov}, \citenamefont {Lucchetti}, \citenamefont {Brown} \emph {et~al.}}]{ryan2021realization}%
  \BibitemOpen
  \bibfield  {author} {\bibinfo {author} {\bibfnamefont {C.}~\bibnamefont {Ryan-Anderson}}, \bibinfo {author} {\bibfnamefont {J.}~\bibnamefont {Bohnet}}, \bibinfo {author} {\bibfnamefont {K.}~\bibnamefont {Lee}}, \bibinfo {author} {\bibfnamefont {D.}~\bibnamefont {Gresh}}, \bibinfo {author} {\bibfnamefont {A.}~\bibnamefont {Hankin}}, \bibinfo {author} {\bibfnamefont {J.}~\bibnamefont {Gaebler}}, \bibinfo {author} {\bibfnamefont {D.}~\bibnamefont {Francois}}, \bibinfo {author} {\bibfnamefont {A.}~\bibnamefont {Chernoguzov}}, \bibinfo {author} {\bibfnamefont {D.}~\bibnamefont {Lucchetti}}, \bibinfo {author} {\bibfnamefont {N.}~\bibnamefont {Brown}}, \emph {et~al.},\ }\bibfield  {title} {\bibinfo {title} {Realization of real-time fault-tolerant quantum error correction},\ }\href {https://doi.org/10.1103/PhysRevX.11.041058} {\bibfield  {journal} {\bibinfo  {journal} {Phys. Rev. X}\ }\textbf {\bibinfo {volume} {11}},\ \bibinfo {pages} {041058} (\bibinfo {year} {2021})}\BibitemShut {NoStop}%
\bibitem [{\citenamefont {Wan}\ \emph {et~al.}(2020)\citenamefont {Wan}, \citenamefont {J{\"o}rdens}, \citenamefont {Erickson}, \citenamefont {Wu}, \citenamefont {Bowler}, \citenamefont {Tan}, \citenamefont {Hou}, \citenamefont {Wineland}, \citenamefont {Wilson},\ and\ \citenamefont {Leibfried}}]{wan2020ion}%
  \BibitemOpen
  \bibfield  {author} {\bibinfo {author} {\bibfnamefont {Y.}~\bibnamefont {Wan}}, \bibinfo {author} {\bibfnamefont {R.}~\bibnamefont {J{\"o}rdens}}, \bibinfo {author} {\bibfnamefont {S.~D.}\ \bibnamefont {Erickson}}, \bibinfo {author} {\bibfnamefont {J.~J.}\ \bibnamefont {Wu}}, \bibinfo {author} {\bibfnamefont {R.}~\bibnamefont {Bowler}}, \bibinfo {author} {\bibfnamefont {T.~R.}\ \bibnamefont {Tan}}, \bibinfo {author} {\bibfnamefont {P.-Y.}\ \bibnamefont {Hou}}, \bibinfo {author} {\bibfnamefont {D.~J.}\ \bibnamefont {Wineland}}, \bibinfo {author} {\bibfnamefont {A.~C.}\ \bibnamefont {Wilson}},\ and\ \bibinfo {author} {\bibfnamefont {D.}~\bibnamefont {Leibfried}},\ }\bibfield  {title} {\bibinfo {title} {Ion transport and reordering in a 2{D} trap array},\ }\href {https://doi.org/10.1002/qute.202000028} {\bibfield  {journal} {\bibinfo  {journal} {Adv. Quantum Technol.}\ }\textbf {\bibinfo {volume} {3}},\ \bibinfo {pages} {2000028} (\bibinfo {year} {2020})}\BibitemShut {NoStop}%
\bibitem [{\citenamefont {Kribs}\ \emph {et~al.}(2005)\citenamefont {Kribs}, \citenamefont {Laflamme},\ and\ \citenamefont {Poulin}}]{kribs2005unified}%
  \BibitemOpen
  \bibfield  {author} {\bibinfo {author} {\bibfnamefont {D.}~\bibnamefont {Kribs}}, \bibinfo {author} {\bibfnamefont {R.}~\bibnamefont {Laflamme}},\ and\ \bibinfo {author} {\bibfnamefont {D.}~\bibnamefont {Poulin}},\ }\bibfield  {title} {\bibinfo {title} {Unified and generalized approach to quantum error correction},\ }\href {https://doi.org/10.1103/PhysRevLett.94.180501} {\bibfield  {journal} {\bibinfo  {journal} {Phys. Rev. Lett.}\ }\textbf {\bibinfo {volume} {94}},\ \bibinfo {pages} {180501} (\bibinfo {year} {2005})}\BibitemShut {NoStop}%
\bibitem [{\citenamefont {Preskill}(1998)}]{preskill1998fault}%
  \BibitemOpen
  \bibfield  {author} {\bibinfo {author} {\bibfnamefont {J.}~\bibnamefont {Preskill}},\ }\href {https://authors.library.caltech.edu/3930/} {\bibinfo {title} {Fault-tolerant quantum computers}} (\bibinfo {year} {1998})\BibitemShut {NoStop}%
\bibitem [{\citenamefont {Qiskit}()}]{Qiskit_quantuminfo}%
  \BibitemOpen
  \bibfield  {author} {\bibinfo {author} {\bibnamefont {Qiskit}},\ }\href@noop {} {\bibinfo {title} {Quantum information}},\ \bibinfo {note} {\url{https://docs.quantum.ibm.com/api/qiskit/quantum_info}}\BibitemShut {NoStop}%
\bibitem [{\citenamefont {Kanazawa}\ \emph {et~al.}(2023)\citenamefont {Kanazawa}, \citenamefont {Egger}, \citenamefont {Ben-Haim}, \citenamefont {Zhang}, \citenamefont {Shanks}, \citenamefont {Aleksandrowicz},\ and\ \citenamefont {Wood}}]{Qiskit_Experiments_A_2023}%
  \BibitemOpen
  \bibfield  {author} {\bibinfo {author} {\bibfnamefont {N.}~\bibnamefont {Kanazawa}}, \bibinfo {author} {\bibfnamefont {D.~J.}\ \bibnamefont {Egger}}, \bibinfo {author} {\bibfnamefont {Y.}~\bibnamefont {Ben-Haim}}, \bibinfo {author} {\bibfnamefont {H.}~\bibnamefont {Zhang}}, \bibinfo {author} {\bibfnamefont {W.~E.}\ \bibnamefont {Shanks}}, \bibinfo {author} {\bibfnamefont {G.}~\bibnamefont {Aleksandrowicz}},\ and\ \bibinfo {author} {\bibfnamefont {C.~J.}\ \bibnamefont {Wood}},\ }\bibfield  {title} {\bibinfo {title} {{Qiskit Experiments: A Python package to characterize and calibrate quantum computers}},\ }\href {https://doi.org/10.21105/joss.05329} {\bibfield  {journal} {\bibinfo  {journal} {Journal of Open Source Software}\ }\textbf {\bibinfo {volume} {8}},\ \bibinfo {pages} {5329} (\bibinfo {year} {2023})}\BibitemShut {NoStop}%
\bibitem [{\citenamefont {Huang}\ \emph {et~al.}(2023{\natexlab{b}})\citenamefont {Huang}, \citenamefont {Brown},\ and\ \citenamefont {Cetina}}]{huang2023comparing}%
  \BibitemOpen
  \bibfield  {author} {\bibinfo {author} {\bibfnamefont {S.}~\bibnamefont {Huang}}, \bibinfo {author} {\bibfnamefont {K.~R.}\ \bibnamefont {Brown}},\ and\ \bibinfo {author} {\bibfnamefont {M.}~\bibnamefont {Cetina}},\ }\bibfield  {title} {\bibinfo {title} {Comparing {S}hor and {S}teane error correction using the {B}acon-{S}hor code},\ }\href {https://doi.org/10.48550/arXiv.2312.10851} {\bibfield  {journal} {\bibinfo  {journal} {arXiv preprint arXiv:2312.10851}\ } (\bibinfo {year} {2023}{\natexlab{b}})}\BibitemShut {NoStop}%
\bibitem [{\citenamefont {Bluvstein}\ \emph {et~al.}(2024)\citenamefont {Bluvstein}, \citenamefont {Evered}, \citenamefont {Geim}, \citenamefont {Li}, \citenamefont {Zhou}, \citenamefont {Manovitz}, \citenamefont {Ebadi}, \citenamefont {Cain}, \citenamefont {Kalinowski}, \citenamefont {Hangleiter} \emph {et~al.}}]{bluvstein2024logical}%
  \BibitemOpen
  \bibfield  {author} {\bibinfo {author} {\bibfnamefont {D.}~\bibnamefont {Bluvstein}}, \bibinfo {author} {\bibfnamefont {S.~J.}\ \bibnamefont {Evered}}, \bibinfo {author} {\bibfnamefont {A.~A.}\ \bibnamefont {Geim}}, \bibinfo {author} {\bibfnamefont {S.~H.}\ \bibnamefont {Li}}, \bibinfo {author} {\bibfnamefont {H.}~\bibnamefont {Zhou}}, \bibinfo {author} {\bibfnamefont {T.}~\bibnamefont {Manovitz}}, \bibinfo {author} {\bibfnamefont {S.}~\bibnamefont {Ebadi}}, \bibinfo {author} {\bibfnamefont {M.}~\bibnamefont {Cain}}, \bibinfo {author} {\bibfnamefont {M.}~\bibnamefont {Kalinowski}}, \bibinfo {author} {\bibfnamefont {D.}~\bibnamefont {Hangleiter}}, \emph {et~al.},\ }\bibfield  {title} {\bibinfo {title} {Logical quantum processor based on reconfigurable atom arrays},\ }\href {https://doi.org/10.1038/s41586-023-06927-3} {\bibfield  {journal} {\bibinfo  {journal} {Nature}\ }\textbf {\bibinfo {volume} {626}},\ \bibinfo {pages} {58} (\bibinfo {year} {2024})}\BibitemShut {NoStop}%
\end{thebibliography}%

\newpage

\appendix
\section{FT code switching between $[[7, 1, 3]]$ and $[[10, 1, 2]]$}\label{app:EC_Codes}

\subsection*{Stabilizer definitions}
The stabilizer generators of the seven-qubit color code $[[7, 1, 3]]$ are given by
\begin{align}
    A_X^{(1)} &= X_1 X_2 X_3 X_4,  \quad A_Z^{(1)} = Z_1 Z_2 Z_3 Z_4 \nonumber\\
    A_X^{(2)} &= X_2 X_3 X_5 X_6 , \quad A_Z^{(2)} = Z_2 Z_3 Z_5 Z_6 \\
    A_X^{(3)} &= X_3 X_4 X_6 X_7,  \quad A_Z^{(3)} = Z_3 Z_4 Z_6 Z_7\nonumber. 
\end{align}

The logical qubit of the $[[10, 1, 2]]$ code is defined by the stabilizer generators
\begin{align}
	B_X^{(1)} &= X_1 X_2 X_3 X_4 X_8 \nonumber \\ 
 B_X^{(2)} &= X_2 X_3 X_5 X_6 X_9  \label{code_10:x_stabs} \\ 
 B_X^{(3)} &= X_3 X_4 X_6 X_7 X_{10} \nonumber 
\end{align}
and
\begin{align}
	B_Z^{(1)} &= Z_1 Z_2 Z_3 Z_4, \quad  B_Z^{(4)} = Z_3 Z_6 Z_8 \nonumber \\
	B_Z^{(2)} &= Z_2 Z_3 Z_5 Z_6, \quad  B_Z^{(5)} = Z_3 Z_4 Z_9 \label{code_10:z_stabs}
\\
	B_Z^{(3)} &= Z_3 Z_4 Z_6 Z_7, \quad  B_Z^{(6)} = Z_2 Z_3 Z_{10}. \nonumber 
\end{align}

For both codes, the logical Pauli operators can be implemented by applying
\begin{align}
    X_L &= X_1 X_2 X_3 X_4 X_5 X_6 X_7,\\ \nonumber
    Z_L &= Z_1 Z_2 Z_3 Z_4 Z_5 Z_6 Z_7. 
\end{align}

\subsection*{Switching operations}
We measure stabilizer operators and apply local Pauli operations that correspond to so-called gauge operators of the subsystem code~\cite{poulin2005stabilizer, kribs2005unified} to switch between the seven-qubit color code $[[7, 1, 3]]$ and the $[[10, 1, 2]]$ code. The lookup tables for switching in both directions are shown in App. Tab.~\ref{tab:lookuptable_switching}. Fig.~\ref{fig:overview_switching_ion_trap} exemplarily illustrates the scheme for switching between $[[10, 1, 2]]$ and $[[7, 1, 3]]$. 

\begin{table}[htb]
    \centering
    \renewcommand*{\arraystretch}{1.3}
    \begin{tabular}{|c|c|}
    \hline
    Measured syndrome & Switching\\
    $(A_X^{(1)}, A_X^{(2)}, A_X^{(3)})$ & operation\\
    \hline
    $(0, 0, 0)$ & $-$\\
    \hline
    $(1, 0, 0)$ & $Z_3 Z_6 Z_8$\\
    \hline
    $(0, 1, 0)$ & $Z_3 Z_4 Z_9$\\
    \hline
    $(0, 0, 1)$ & $Z_2 Z_3 Z_{10}$\\
    \hline
    $(1, 1, 0)$ & $Z_4 Z_6 Z_8 Z_9$\\
    \hline
    $(1, 0, 1)$ & $ Z_2 Z_6 Z_8 Z_{10}$\\
    \hline
    $(0, 1, 1)$ & $ Z_2 Z_4 Z_9 Z_{10}$\\
    \hline
    $(1, 1, 1)$ & $ Z_2 Z_3 Z_4 Z_6 Z_8 Z_9 Z_{10}$\\
    \hline
    \hline
    Measured syndrome & Switching \\
    $(B_Z^{(4)}, B_Z^{(5)}, B_Z^{(6)})$ & operation\\
    \hline
    $(0, 0, 0)$ & $-$ \\
    \hline
    $(1, 0, 0)$ & $X_1 X_2 X_3 X_4$\\
    \hline
    $(0, 1, 0)$ & $X_2 X_3 X_5 X_6$\\
    \hline
    $(0, 0, 1)$ & $X_3 X_4 X_6 X_7$\\
    \hline
    $(1, 1, 0)$ & $X_1 X_4 X_5 X_6$\\
    \hline
    $(1, 0, 1)$ & $X_1 X_2 X_6 X_7$\\
    \hline
    $(0, 1, 1)$ & $X_2 X_4 X_5 X_7$\\
    \hline
    $(1, 1, 1)$ & $X_1 X_3 X_5 X_7$\\
    \hline
    \end{tabular}
    \caption{\justifying \textbf{Lookup tables for switching between $[[7, 1, 3]]$ and $[[10, 1, 2]]$. }For switching from $[[10, 1, 2]]$ to $[[7, 1, 3]]$ (top) we measure the three $X$-stabilizers $(A_X^{(1)}, A_X^{(2)}, A_X^{(3)})$ and apply a Pauli $Z$-operation that fixes the state into the codespace of the seven-qubit color code while preserving the encoded information. For the inverse direction (bottom), we measure the three weight-3 $Z$-stabilizers of the $[[10, 1, 2]]$ code $(B_Z^{(4)}, B_Z^{(5)}, B_Z^{(6)})$ and apply a suitable Pauli $X$-operation. }
    \label{tab:lookuptable_switching}
\end{table}

\subsection*{FT switching protocols}
\begin{figure*}
    \centering
    \includegraphics[width=180mm]{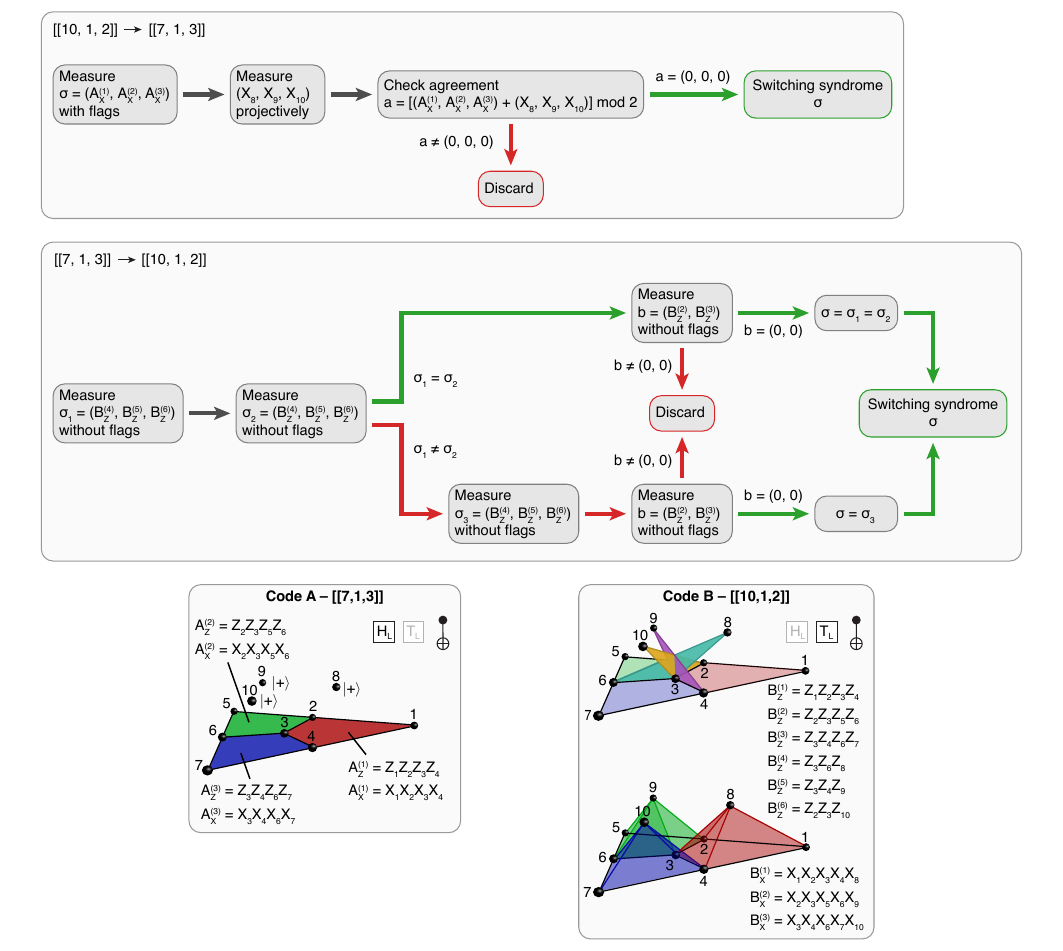}
    \caption{\justifying \textbf{Protocols for switching between [[7, 1, 3]] and [[10, 1, 2]] codes.} Stabilizers of the target code are measured to obtain the switching syndrome $\sigma$. If an error happens (indicated with red arrows), the run must be discarded or additional measurements must be taken.}
    \label{fig:protocol}
\end{figure*}

The following two algorithms summarize the protocols for FT switching between $[[10, 1, 2]]$ and $[[7, 1, 3]]$, which are described in Sec.~\ref{sec:theory}. The graphical representation of the protocols is also given in Fig.~\ref{fig:protocol}.

\begin{table}[htb]
    \centering
    \renewcommand*{\arraystretch}{1.3}
    \begin{tabular}{l}
    \hline
         \textbf{Protocol}: FT switching $[[10, 1, 2]] \rightarrow [[7, 1, 3]]$\\
    \hline
    \textbf{Input: }Logical state in $[[10, 1, 2]]$\\
    \textbf{Output: }Logical state in $[[7, 1, 3]]$\\
    1:  Measure $ \sigma = (A^{(1)}_X$, $A^{(2)}_X$, $A^{(3)}_X)$ with flags\\
    2: \quad \quad \quad If a circuit flags: discard\\
    3: Measure qubits 8, 9 and 10 in the $X$-basis\\
    4: Check agreement of opposing $X$-operators\\
    5: \quad  \quad \quad$a = [(A^{(1)}_X$, $A^{(2)}_X$, $A^{(3)}_X) + (X_8, X_9, X_{10})]  \mod  2$\\
    6: \quad  \quad  \quad If $a \neq $(0, 0, 0): discard \\
    7: Apply switching operation according to lookup table \\ 
    \;\quad using switching syndrome $\sigma$ \\
    \hline
    \end{tabular}
    \label{tab:protocol_summary_10_to_7}
\end{table}

\begin{table}[tb]
    \centering
    \renewcommand*{\arraystretch}{1.3}
    \begin{tabular}{l}
    \hline
         \textbf{Protocol}: FT switching $[[7, 1, 3]] \rightarrow [[10, 1, 2]]$\\
    \hline
    \textbf{Input: }Logical state in $[[7, 1, 3]]$\\
    \textbf{Output: }Logical state in $[[10, 1, 2]]$\\
    1:\;\: Measure $\sigma_1 = (B^{(4)}_Z$, $B^{(5)}_Z$, $B^{(6)}_Z)$\\
    2:\;\: Measure $\sigma_2 = (B^{(4)}_Z$, $B^{(5)}_Z$, $B^{(6)}_Z)$\\
    3a:  If 1 and 2 agree:  \\
    4a:  \quad \quad \quad Measure $(B^{(2)}_Z$, $B^{(3)}_Z$) \\
    5a:  \quad \quad \quad If $(B^{(2)}_Z, B^{(3)}_Z) \neq$ (0, 0): discard \\
    6a: \quad \quad \quad $\sigma_{\mathrm{final}} = \sigma_2$ \\
    3b:  If 1 and 2 disagree: \\
    4b: \quad \quad \quad Measure $\sigma_3 = (B^{(4)}_Z$, $B^{(5)}_Z$, $B^{(6)}_Z)$ and $(B^{(2)}_Z, B^{(3)}_Z)$ \\
    5b: \quad \quad \quad  If $(B^{(2)}_Z, B^{(3)}_Z) \neq$ (0, 0): discard \\
    6b: \quad \quad \quad $\sigma_{\mathrm{final}} = \sigma_3$ \\
    7:\;\: Apply switching operation according to lookup table \\ \quad ~ using switching syndrome $\sigma_{final}$ \\
    \hline
    \end{tabular}
    \label{tab:protocol_summary_7_to_10}
\end{table}

Stabilizers can be measured by coupling a physical auxiliary qubit to the data qubits that belong to the operator to be measured~\cite{preskill1998fault}. However, this scheme is not FT because single faults on auxiliary qubits can directly result in a logical failure. This is avoided by making use of flag-qubits~\cite{goto2016minimizing, chao2018quantum, chamberland2019fault}.
In addition, we also need to be able to identify errors on data qubits that invert the projective measurement with random outcomes~\cite{butt2023fault}. For example, consider the case where a $Z$-error occurs on qubit 1: If we originally would have directly (randomly) projected onto the correct target codespace and measured the trivial switching syndrome (0, 0, 0), we would now measure the same one as illustrated in Fig.~\ref{fig:overview_switching_ion_trap}. This would cause us to apply the same gauge operator $Z_3 Z_6 Z_8$ as before, which in total amounts to a logical $Z_L = Z_1 Z_3 Z_6$ on the target Steane code. 
There are different strategies for detecting these dangerous errors on data qubits for each switching direction. For switching from $[[7, 1, 3]]$ to $[[10, 1, 2]]$, we identify the potentially dangerous positions and perform additional stabilizer measurements to check if an error has occurred at one of these. For the inverse direction, we can detect these dangerous errors on data qubits without additional stabilizer measurements. After switching in this direction, we measure qubits 8, 9, and 10 in the $X$-basis and compare this outcome to their opposing stabilizer plaquette. For example, we compare the outcome of qubit 8 to the outcome of $A_X^{(1)} = X_1 X_2 X_3 X_4$. Since we started in a +1-eigenstate of the weight-5 cell, we know that pairs of these opposing operators have to agree so that, in total, they amount to $0 \mod 2$.

\section{Noise model and simulation methods}\label{app:numerics}

We determine the expected fidelities of the implemented code switching protocols by performing Monte Carlo (MC) simulations. In the numerical simulations, every ideal circuit component is followed by an error $E$ with a specified probability $p$, that is in particular the probability of an error occurring $p\in \left[0,1 \right]$. We include a depolarizing channel on all single- and two-qubit gates, which is defined by the error rates $p_1$ and $p_2$ with which one of the errors in the error sets $E_1$ and $E_2$ is applied. The error channel is defined as 
\begin{align}
    \mathcal{E}_1(\rho) &= (1 - p_1)\rho + \frac{p_1}{3} \sum_{i= 1}^3 E^{i}_1 \rho E^{i}_1 \\
    \mathcal{E}_2(\rho) &= (1 - p_2)\rho + \frac{p_2}{15} \sum_{i= 1}^{15}   E_2^{i} \, \rho\, E_2^{i}.  \nonumber\label{eq:depol_single_qubit}
\end{align}
with the error sets $E_1 \in \{X$, $Y$, $Z\}$ for  $k = 1, 2, 3$ and 
$E_2$ $\in$ $\{IX$, $XI$, $XX$, $IY$, $YI$, $YY$, $IZ$, $ZI$, $ZZ$, $XY$, $YX$, $XZ$, $ZX$, $YZ$, $ZY\}$ for  $k = 1, ..., 15$. 
Qubits are initialized and measured in the computational basis and faults on these two operations are simulated by applying $X$-flips after and before the respective operation with probabilities $p_{\mathrm{init}}$ and $p_{\mathrm{meas}}$. Furthermore, idling qubits experience dephasing due to environmental fluctuations. 
We model this dephasing of idling qubits with the error channel 
\begin{align}
    \mathcal{E}_{\mathrm{idle}}(\rho) &= (1 - p_{\mathrm{idle}})\rho + p_{\mathrm{idle}} Z\rho Z. 
\end{align}
Here, a $Z$-fault is placed on each idling qubit with probability $p_{\mathrm{idle}}$, which depends on the execution time $t$ of the performed gate and the coherence time $T_2 \approx 50\,$ms 
\begin{align}
    p_{\mathrm{idle}} = \frac{1}{2} \left[1 - \mathrm{exp}\left(-\frac{t}{T_2} \right) \right]. 
\end{align}
Lastly, we perform mid-circuit detections, where auxiliary qubits are measured while the data qubits are kept intact, as discussed in Sec.~\ref{sec:experimental_setup}. Based on single qubit process tomography, we estimate the error rates of $X$- $Y$- and $Z$-faults on the idling data qubits~\cite{postler2023demonstration} and model the mid-circuit detection with an asymmetric depolarizing channel which is specified by individual Pauli $p^{   (x)}_{\mathrm{inseq}}, p^{(y)}_{\mathrm{inseq}}, p^{(z)}_{\mathrm{inseq}}$ error rates. All error rates and relevant gate durations are summarized in App. Tab.~\ref{tab:error_rates_simulation}. 

\begin{table}[tb]
    \centering
    \renewcommand*{\arraystretch}{1.3}
    \begin{tabular}{|c|c|c|}
        \hline
         Operation & Error rate & Duration\\
         \hline
         Two-qubit gate & $p_{2} = 0.027$ & \SI{322.5}{\micro s}\\
         \hline
         Single-qubit gate & $p_{1} = 0.0036$ & \SI{25}{\micro s}\\
         \hline
         Measurement & $p_{\mathrm{meas}} = 0.003$& \\
         \hline
         Preparation & $p_{\mathrm{init}} = 0.003$& \\
         \hline
         & $p^{(x)}_{\mathrm{mid-circ}}= 0.011$ & \\ 
         Mid-circuit detection  & $p^{(y)}_{\mathrm{mid-circ}}= 0.024$ & \\ 
         & $p^{(z)}_{\mathrm{mid-circ}}= 0.035$ & \\
         \hline
                  \hline

         Coherence time &  \multicolumn{2}{c|}{$T_2 = \SI{50}{\milli\second}$}  \\
        \hline

    \end{tabular}
    \caption{\justifying \textbf{Depolarizing error rates and duration of operations on a trapped-ion quantum processor~\cite{postler2023demonstration}. }
    }
    \label{tab:error_rates_simulation}
\end{table}

We determine the state fidelity between two states $\rho_1$ and $\rho_2$ as
\begin{align}
    F(\rho_1, \rho_2) = \mathrm{Tr}{\left[\sqrt{\sqrt{\rho_1} \rho_2 \sqrt{\rho_1} } \right]^2}.
\end{align}
 The process fidelity is calculated similarly with
\begin{align}
    F_{\mathrm{pro}}(\mathcal{E, F}) = F(\rho_{\mathcal{E}}, \rho_{\mathcal{F}}),
\end{align}
where $F$ is the state fidelity, and $\rho_{\mathcal{E}}$ the normalized Choi matrix for channel $\mathcal{E}$. We use Qiskit's Quantum Information package to calculate fidelities~\cite{Qiskit_quantuminfo}.

\section{Experimental methods}\label{app:experiment}

\subsection{Mid-circuit measurement}

\begin{figure*}[tb]
	\centering
	\includegraphics[width=120mm]{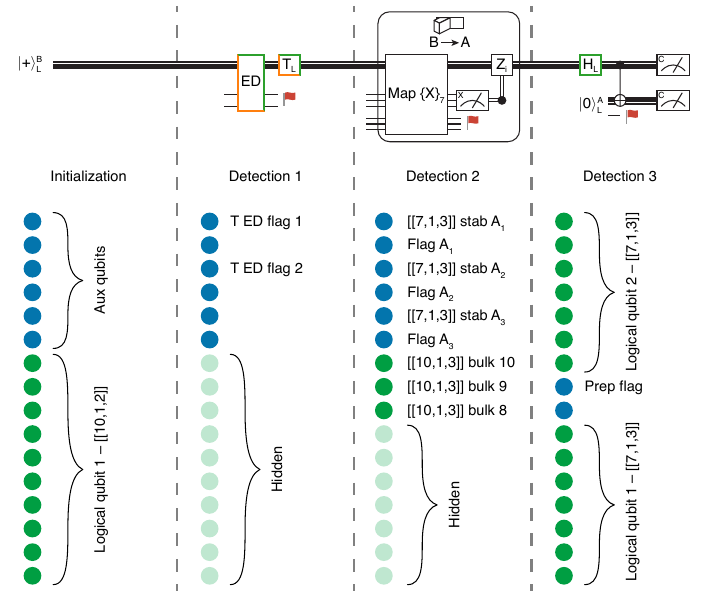}
	\caption{\justifying \textbf{Auxiliary qubits mapping for the entangled states generation circuit.} 
Data qubits encoding logical states are shown in green, while auxiliary qubits used for stabilizers/flags mapping are shown in blue. There are three detections happening during the circuit, two of which are mid-circuit detections. All auxiliary qubits are detected during mid-circuit detections while some data qubits are hidden (shown in light green). The information mapped to auxiliary qubits during each detection is shown next to qubits, no label means that the qubit is not used but still detected. After any mid-circuit detection all detected qubits are reinitialized as $\ket{0}$. All the qubits are detected during the last detection. 
}
	\label{fig:cnot_detailed}
\end{figure*}

The mid-circuit measurement in our implementation setup is a significant technical source of errors with the current parameters (see App. Tab.~\ref{tab:error_rates_simulation}). Therefore, we aim at constructing our circuits in a way that minimizes the required number of mid-circuit detections. Stabilizers and flags mapped to auxiliary qubits are measured all in one mid-circuit measurement whenever possible, an example of which is shown in Fig.~\ref{fig:cnot_detailed}. 
When not all of the auxiliary qubits are used we aim at mapping information to the auxiliary qubits spatially far away from the data qubits to avoid optical cross-talk on data qubits. It is also beneficial to leave idle buffer qubits between auxiliary qubits to avoid cross-talk between auxiliary qubits. However, minimizing the number of mid-circuit measurements takes priority. 

We specify the number of mid-circuit measurements for each protocol presented here in App. Tab.~\ref{tab:n_detections}.

\begin{table}[]
\begin{tabular}{l|c|c|}
\cline{2-3}
                                                     & TQ gates    & Mid-circ         \\ \hline
\multicolumn{1}{|l|}{$|0\rangle_L^{B} $ } & 15        & 0                   \\ \hline
\multicolumn{1}{|l|}{$|+\rangle_L^{B} $ } & 13       & 0                   \\ \hline
\multicolumn{1}{|l|}{T$_L$}                          & 12        & 0                   \\ \hline
\multicolumn{1}{|l|}{$|\psi\rangle_L^{B} \rightarrow |\psi\rangle_L^{A} $} & 18& 0\\ \hline
\multicolumn{1}{|l|}{$|\psi\rangle_L^{A} \rightarrow |\psi\rangle_L^{B}$}  & 26/35& 1/2\\ \hline
\multicolumn{1}{|l|}{T$_L |+\rangle_L^{B} \rightarrow $T$_L |+\rangle_L^{A}$ }& 43 & 1\\ \hline
\multicolumn{1}{|l|}{Clifford CNOT-protocol}         & 49        & 1                     \\ \hline
\multicolumn{1}{|l|}{non-Clifford CNOT-protocol}     & 61        & 2                     \\ \hline
\end{tabular}
    \caption{\justifying \textbf{Number of two-qubit gates and mid-circuit measurements for different FT protocols.} Code B corresponds to $[[10, 1, 2]]$ and code A to $[[7, 1, 3]]$. The CNOT-protocol includes the FT initialization of $|+\rangle_L^{B}$, FT switching to the seven-qubit color code, the initialization of a second logical qubit $|0\rangle_L^{A} $ and a transversal CNOT-gate, and has the highest number of two-qubit gates and mid-circuits measurements. }
    \label{tab:n_detections}
\end{table}

\subsection{Flag bunching}
FT code switching $[[10, 1, 2]]$ $\rightarrow$ $[[7, 1, 3]]$ includes the measurement of three stabilizers with flags, which requires six auxiliary qubits in total. 
Preparing the logical $\ket{0}_L$ state of the $[[10, 1, 2]]$ also requires one auxiliary qubit as a flag for fault-tolerance. Therefore, seven auxiliary qubits are required to map all the required stabilizers and flags if $[[10, 1, 2]] \rightarrow [[7, 1, 3]]$ code switching is done immediately after the state preparation, which is the case for building block characterization (see Fig.~\ref{fig:blocks}, App. Tab.~\ref{tab:fidelities_building_blocks}). In the current experimental configuration we are limited to no more than six auxiliary qubits in addition to 10 data qubits, hence we map two flags to the same auxiliary qubits to reduce the number of mid-circuit measurements. Bunching together the preparation flag and one of the stabilizer flags still preserves fault tolerance in this particular case: if only one error takes place, no dangerous error propagation can happen in such a way that both flags should be raised at the same time. Such flag bunching only takes place in this protocol.

\subsection{Branching post-selection}
FT code switching $[[7, 1, 3]]$ $\rightarrow$ $[[10, 1, 2]]$ requires mid-circuit decision making to choose which set of stabilizers should be measured based on the result of the first mid-circuit measurement (see step 3a,b in the description of the protocol in App.~\ref{app:EC_Codes}). The current hardware electronics configuration does not allow for fast communication between the camera and the control electronics so mid-circuit decision making within the qubits' lifetime is currently not feasible. Thus, we assume we know the result of the measurement and act accordingly instead of making a decision based on the result of the mid-circuit measurement. We discard this experimental shot if during the analysis it turns out that our assumption about the measurement result was incorrect. This effectively decreases the acceptance rate for $[[7, 1, 3]]$ $\rightarrow$ $[[10, 1, 2]]$ by a factor of two compared to the protocol with mid-circuit decision making. 

\subsection{Switching operation}

Both code switching procedures require the application of a switching operation based on the outcome of the measurement of the stabilizers of the target code (see App. Tab.~\ref{tab:lookuptable_switching}). This switching operation cannot be applied mid-circuit since the setup is currently missing a mid-circuit decision-making feature. Therefore, we apply switching operations in classical processing with Pauli frame updates. This is possible if all the gates after the switching operation belong to the Clifford group, which is the case for all of our circuits. 

\subsection{Number of measurements}
The total numbers of measurements performed for different protocols before post-selection are the following. Building blocks: 30000 shots per measurement basis for each input state for [[7,1,3]] $\rightarrow$ [[10,1,2]] switching protocol and 12500 shots for the other protocols. Bloch sphere states: 5000 shots per measurement basis for each prepared state. CNOT protocol: 15000 shots per measurement basis for each configuration of the protocol.

\section{FT code switching building blocks}\label{app:FT_building_blocks}

App. Tab.~\ref{tab:fidelities_building_blocks} summarizes the obtained process fidelities for all code switching building blocks that we implemented experimentally and numerically. 

We observe that the fidelity for nFT switching from $[[7, 1, 3]]$ to $[[10, 1, 2]]$ is higher than for FT switching. This has two main reasons: The first is that for the nFT protocol, we only measure weight-3 $Z$-stabilizers. This means that there are no dangerous faults on auxiliary qubits that could result in a logical error. Any error propagating from the auxiliary qubits to the data qubits is only equivalent to a weight-1 $Z$-error and at least detectable on the target code. The nFT protocol therefore only requires nine two-qubits gates and there are few possible single-fault positions on data qubits that can cause a logical failure~\cite{butt2023fault}. Furthermore, we require almost twice as many two-qubit gates for the FT implementation of this direction than for the inverse one and have to perform a mid-circuit measurement, as summarized in App. Tab.~\ref{tab:n_detections}. This overhead in noisy operations in the FT protocol leads to a decrease in fidelity with the current level of noise in gate operations.

\begin{table*}[tb] 
    \centering
    \begin{tabular}{| l | c | c | c | c | c | c |}
        \cline{2-7}
         \multicolumn{1}{c|}{} & \multicolumn{3}{c|}{Experiment} & \multicolumn{3}{c|}{Simulation} \\
        \multicolumn{1}{c|}{} & nFT & FT & AR (\%) & nFT & FT & AR (\%) \\
        \hline
        $|\psi\rangle_L^B$ & 0.963(11) & 0.965(8)  & 86 / 78 & 0.965(3)  & 0.959(3) & 85 / 76\\
        \hline
        $T_L |\psi\rangle_L^B$ & 0.761(27) & 0.755(14)  & 81 / 51 &  0.777(5) & 0.780(6) &  81 / 49\\
        \hline
        $|\psi\rangle_L^B \rightarrow |\psi\rangle_L^A$ & 0.586(9) & 0.877(15) & 100 / 23 & 0.576(4) & 0.903(7) & 100 / 27\\
        \hline
        $|\psi\rangle_L^A \rightarrow |\psi\rangle_L^B$  & 0.832(19) & 0.612(16) & 79 / 14 & 0.832(3) & 0.643(5) & 80 / 31\\
        \hline
    \end{tabular}
    \caption{\justifying \textbf{Process fidelities and acceptance rates of code switching building blocks.} Code A refers to the seven-qubit color code $[[7, 1, 3]]$ and code B to the 10-qubit code $[[10, 1, 2]]$. We determine the fidelities for each code switching building block experimentally and numerically for the nFT as well as the FT protocol version using the methods described in Apps.~\ref{app:numerics} and~\ref{app:experiment}. The numbers in brackets indicate the uncertainty on the obtained value of the fidelity, which are determined as discussed in App.~\ref{tomo_details}. The acceptance rates (AR) are given for the nFT and the FT protocols, as for example for the initialization of a logical states on code B, 86\% of the runs are accepted in the nFT case. 
    }
    \label{tab:fidelities_building_blocks}
    \end{table*}

\begin{figure}[ht]
	\centering
	\includegraphics[width=89mm]{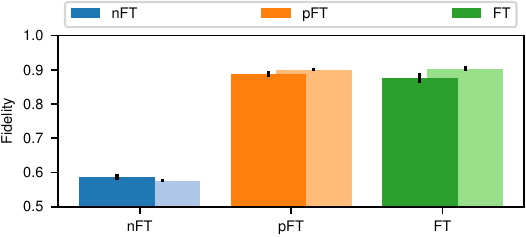}
	\caption{\justifying \textbf{Fidelities for partially FT code switching. } 
    Logical process fidelities of switching from the $[[10, 1, 2]]$ to the $[[7, 1, 3]]$ code, using a non-FT (blue), a partially FT (orange) and FT (green) scheme. Partially FT switching does not require additional measurements or auxiliary qubits and only includes the agreement check of opposing pairs of operators. For FT switching, we add flag qubits for fully FT stabilizer measurements. }
	\label{fig:pft}
\end{figure}

For switching from $[[10, 1, 2]]$ to the $[[7, 1, 3]]$ code, the FT scheme consists of two features: first, the flag-based stabilizer measurement schemes and, second, the agreement check of opposing pairs of operators, as discussed in Sec.~\ref{sec:theory}~\cite{butt2023fault}. The latter does not require any additional qubits or measurements, since it can be done completely in classical post-processing. 
We estimate the contributions of these two features to the observed increase in fidelity by performing a partially FT experiment (pFT): we only add the agreement check to the nFT scheme and again, determine the fidelity, which is shown in Fig.~\ref{fig:pft}. We find that the agreement check contributes significantly more to the infidelity than adding flag qubits for FT stabilizer measurements. 

\section{Quantum state and process tomography}
\label{tomo_details}
The numbers given in Sec.~\ref{sec:results} were obtained with logical quantum state/process tomography via linear inversion using the Qiskit Experiments package~\cite{Qiskit_Experiments_A_2023}.
For each protocol in Figs.~\ref{fig:blocks},~\ref{fig:cnot}\textbf{b} all data was collected in five experimental runs. Experimental runs were performed on different days and the setup's performance differed slightly. The results' spread caused by the difference in the setup's performance is more substantial than the one coming from the finite number of shots taken for each protocol. Hence, to estimate errors we performed logical quantum state/process tomography for each experimental run separately and then averaged the results. 
The resulting values are mean and standard deviation over fidelities obtained from each run.
For Fig.~\ref{fig:bloch}\textbf{c} logical quantum state tomography was performed for each state from Fig.~\ref{fig:bloch}\textbf{b} and the results were averaged over the states belonging to the same group (color).
The data from five experimental runs was joined to obtain the average density matrices in Fig.~\ref{fig:cnot}\textbf{c}, which were reconstructed by means of logical quantum state tomography. The reconstructed matrices are the following:

\begin{widetext}   
\begin{equation}
\mathrm{CNOT}_L (\ket{+}_L \otimes \ket{0}_L) = 
\begin{pmatrix}
0.447 &   -0.003-0.002 i & -0.005-0.002 i &  0.231+0.022 i \\
-0.003+0.002 i & 0.057 & 0.007+0.007 i & 0.001  \\
-0.005+0.002 i & 0.007-0.007 i & 0.064 & -0.001 i \\
0.231-0.022 i & 0.001  & 0.001 i & 0.432 
\end{pmatrix}, \nonumber 
\end{equation}

\begin{equation}
\mathrm{CNOT}_L (T_L\ket{+}_L \otimes \ket{0}_L) = 
\begin{pmatrix}
0.429 &  -0.005-0.003 i &  0.003+0.003 i &  0.117-0.101 i \\
-0.005+0.003 i & 0.07  & -0.01 +0.01i &  0.004+0.002 i \\ 
0.003-0.003 i & -0.01 -0.01 i &   0.074 & -0.001-0.002 i \\
0.117+0.101 i & 0.004-0.002 i & -0.001+0.002 i & 0.427 
\end{pmatrix}, \nonumber
\end{equation}

\begin{equation}
\mathrm{CNOT}_L (H_LT_L\ket{+}_L \otimes \ket{0}_L) = 
\begin{pmatrix}
0.477 & 0.029-0.035 i & 0.025-0.035 i & 0.125-0.009i \\
0.029+0.035 i & 0.07 & 0.012-0.006 i & 0.03 -0.018i \\
0.025+0.035 i & 0.012+0.006 i & 0.081 & 0.023-0.013 i \\
0.125+0.009 i & 0.03 +0.018 i & 0.023+0.013 i & 0.373 
\end{pmatrix}.\nonumber
\end{equation}

\end{widetext}

\section{Additional post-selection}
\label{trivial_syndrome}

The logical fidelity of the output state of the circuits can be additionally boosted without any extra measurements or operations by post-selecting for the trivial syndrome~\cite{huang2023comparing, bluvstein2024logical}. The final measurement of the data qubits in every circuit yields stabilizer values for the corresponding QEC code. If all results with non-trivial stabilizer syndromes are discarded, the logical fidelity increases while the acceptance rate decreases. In doing so, a fraction of runs with error configurations of weight $>$1 is sorted out. The resulting logical fidelities and acceptance rates with additional post-selection on trivial stabilizer syndromes for some protocols are given in App. Tab.~\ref{tab:trivial_syndrome}. 

\begin{table*}[]
\begin{tabular}{l|c|c|c|c|}
\cline{2-5}
                            & \multicolumn{2}{c|}{no PS} & \multicolumn{2}{c|}{PS} \\

                                                     & Fidelity    & AR (\%) & Fidelity & AR (\%)       \\ \hline
\multicolumn{1}{|l|}{$|\psi\rangle_L^A \rightarrow |\psi\rangle_L^B$}      & 0.597(15) & 14 & 0.75(4)    & 4    \\ \hline
\multicolumn{1}{|l|}{T$_L |+\rangle_L^B \rightarrow $T$_L |+\rangle_L^A$ } & 0.887(10) & 27 & 0.963(4)   & 19  \\
\multicolumn{1}{|l|}{}                                                                              & 0.818(10) & 15 & 0.919(11)  & 10    \\ 
\multicolumn{1}{|l|}{}                                                                              & 0.890(13) & 27 & 0.959(9)   & 19     \\ 
\multicolumn{1}{|l|}{}                                                                              & 0.810(18) & 12 & 0.918(17)  & 8    \\ \hline

\multicolumn{1}{|l|}{CNOT-protocol}                                                                 & 0.672(16) & 22 & 0.934(7)  & 6    \\ 
\multicolumn{1}{|l|}{}                                                                              & 0.579(16) & 10 & 0.795(15)  & 3    \\ 
\multicolumn{1}{|l|}{}                                                                              & 0.551(27) & 10 & 0.80(5)  & 2   \\ \hline

\end{tabular}
    \caption{\justifying \textbf{Fidelities without and with additional post-selection.} Code A refers to the seven-qubit color code $[[7, 1, 3]]$ and code B to the 10-qubit code $[[10, 1, 2]]$. Additional post-selection (PS) on trivial stabilizer syndrome can be done to boost the logical fidelity at the cost of the acceptance rate. The values for logical state/process fidelities and acceptance rates (AR) with additional post-selection are given for three protocols, as presented in Fig.~\ref{fig:cnot}. Different rows within one cell refer to different states prepared.}
    \label{tab:trivial_syndrome}
\end{table*}

\section{Error budget and rotated $[[10, 1, 2]]$}\label{app:error_budget}

\subsection{Individual error contributions}
\begin{figure*}[tb]
    \centering
    \includegraphics[
    width=120mm
    ]{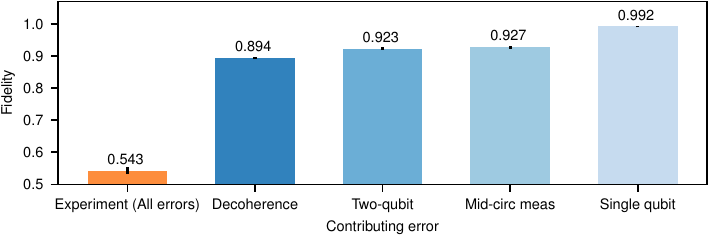}
    \caption{\justifying \textbf{Estimated isolated error contributions to the total infidelity.}
    Fidelities for the non-Clifford CNOT-protocol obtained with numerical simulations. The orange column shows the fidelity for the system with realistic experimental error rates from App. Tab.~\ref{tab:error_rates_simulation}. The blue columns show the results for the reduced error model where all the errors were set to zero except for one with the corresponding fidelity indicated on top of the columns. The non-zero error source varies between blue columns and is specified below. \textit{Decoherence} corresponds to the dephasing of idling qubits while gates are carried out on different sets of qubits, \textit{two-qubit} to the physical two-qubit MS-gate error rate, \textit{mid-circ meas} to the idling of data qubits during the measurement of auxiliary qubits and \textit{single-qubit} to all single-qubit noise processes, including faulty initializations, measurements as well as single-qubit gates.}
    \label{fig:error_budget}
\end{figure*}

We perform numerical simulations of the non-Clifford CNOT protocol with a reduced error model where we set to zero all error sources except for one in order to estimate and compare contributions to the logical infidelity from the different individual error sources present, as shown in Fig~\ref{fig:error_budget}. 
Note that the single contributions are not additive, as is expected, since errors propagate and the different noise processes influence each other. 
Decoherence and mid-circuit measurements have a significant impact on the fidelity, which is comparable to that of two-qubit gate errors. While improving two-qubit gate fidelities in a large ion chain is a complex problem, we anticipate improved coherence and mid-circuit measurement performance pending near-term hardware changes. 

\subsection{Impact of two-qubit entangling gate errors}

\begin{figure*}[!tb]
    \centering
    \includegraphics[width=1\linewidth]{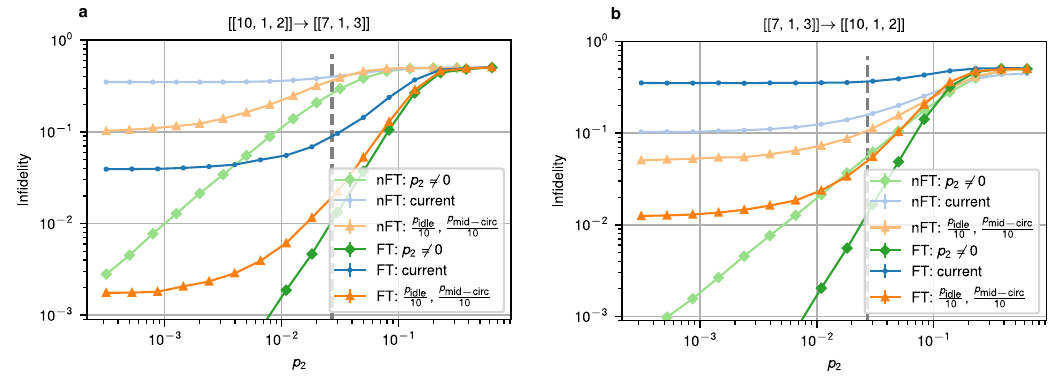}
    \caption{\justifying \textbf{Projected performance of FT code switching between $[[10, 1, 2]]$ and $[[7, 1, 3]]$. } We scale the two-qubit error rate $p_2$ for three different scenarios: first, for the current set of noise parameters (blue), second, for a modified set of parameters, where only the idling dephasing rate $p_{\mathrm{idle}}$ and the mid-circuit detection error rate $p_{\mathrm{mid-circ}}$ is reduced by a factor of 10 (orange) and, third, setting all error rates to 0, except for the two-qubit error rate $p_2$ (green). We consider the FT and nFT switching protocols for switching (a) from $[[10, 1, 2]]$ to $[[7, 1, 3]]$ and (b) in the inverse direction. The grey dashed line corresponds to the current value of $p_2$. }
    \label{fig:projected_performance}
\end{figure*}

We investigate the impact of the different noise processes further by simulating the scaling of the logical infidelity as a function of the two-qubit error rate $p_2$. We consider both nFT and FT switching in both directions for three different settings of the noise parameters to determine how an improvement of the two-qubit error rate would affect the total infidelity as shown in Fig.~\ref{fig:projected_performance}. 
Already for the current noise parameters, FT switching from $[[10, 1, 2]]$ to $[[7, 1, 3]]$ outperforms the nFT protocol, but the infidelity quickly reaches a regime where there is no qualitative change with further improvement of $p_2$. This is in stark contrast to the behavior found with magic state injection~\cite{heussen2023strategies}. 
However, this changes if the dephasing error rate on idling qubits $p_{\mathrm{idle}}$ and on the mid-circuit detections $p_{\mathrm{mid-circ}}$ is reduced tenfold. In this case, the infidelity decreases by more than an order of magnitude, even for small values of $p_2$. 
When switching from $[[7, 1, 3]]$ to $[[10, 1, 2]]$, nFT switching outperforms the FT protocol for the complete range of considered values of $p_2$. This is inverted for a tenfold improvement of $p_{\mathrm{idle}}$ and $p_{\mathrm{mid-circ}}$. 
FT switching now achieves smaller infidelities than the nFT scheme, even for the current value of $p_2$. 
Note that these reduced values for the error rates are within reach as extended coherence times have been demonstrated~\cite{harty2014high, ruster2016long, wang2021single} and composite pulse sequences have been shown to be more robust against crosstalk and laser amplitude noise~\cite{wimperis1989composite,wimperis1994broadband}, which can be applied in the mid-circuit measurements. 

\subsection{Rotated more dephasing-robust encoding}

It has been shown that QEC protocols tailored to experimental systems with biased noise can improve performance on these setups~\cite{bonilla2021xzzx, xu2023tailored, huang2023tailoring, pal2022relaxation}. 
Due to dephasing of idling qubits and during mid-circuit detections, the noise in our experimental setup is strongly $Z$-biased. 
We can reduce the sensitivity to dephasing by simply interchanging the $X$- and $Z$-type stabilizers of the initial $[[10, 1, 2]]$ code, thereby exploiting EC properties which are advantageous given strongly $Z$-biased noise. 
The $[[10, 1, 2]]$ code as described in App.~\ref{app:EC_Codes} is capable of correcting up to three $X$-errors but can only detect at most one $Z$-error. This is an unfortunate combination for the current experimental setup since decoherence of our qubits is one of the dominant error sources. However, the $[[10, 1, 2]]$ can be modified by exchanging the support of $X$- and $Z$-stabilizers (\ref{code_10:x_stabs}), (\ref{code_10:z_stabs}). We refer to the initial code as $[[10, 1, 2]]_Z$ and to the modified one as $[[10, 1, 2]]_X$ or the rotated $[[10, 1, 2]]$ code. The rotated code has its $X$/$Z$ error-correcting properties exchanged, i.e. it is capable of correcting up to three $Z$-errors, but can only detect at most one $X$-error. 
By effectively exchanging the support of the $X$- and $Z$-stabilizers of the $[[10, 1, 2]]_Z$ code, the logical $T$-gate now becomes a rotation about the $X$-axis instead of the $Z$-axis and the physically executed operation includes additional $H$-gates before and after the previous $T$-gate
\begin{equation*}
    T^X_L = H^{\otimes 10} T^Z_L H^{\otimes 10}. 
\end{equation*}

\begin{figure*}[tb]
    \centering
    \includegraphics[width=180mm]{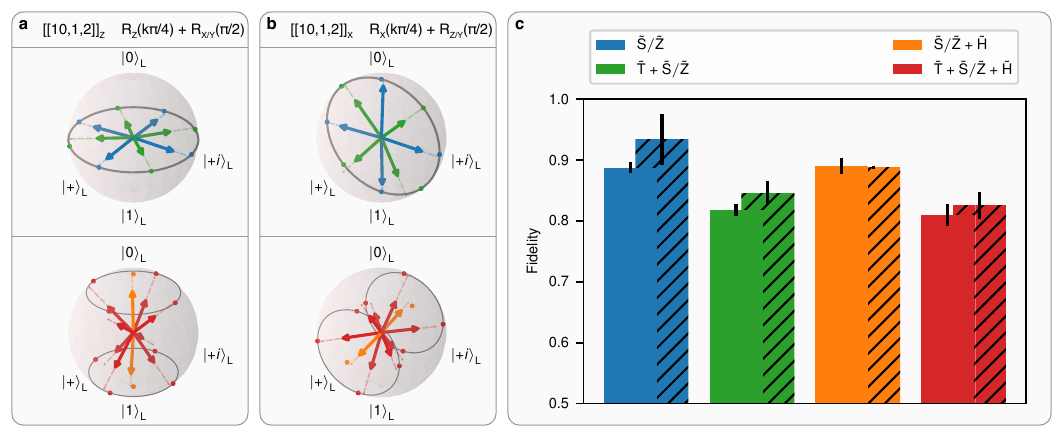}
    \caption{\justifying \textbf{Bloch states and fidelities for the rotated $[[10, 1, 2]]$ code. } (a) Bloch states that were obtained for the original $[[10, 1, 2]]$ code, as shown in Fig.~\ref{fig:bloch}. (b) States within the Bloch sphere that were prepared with the rotated $[[10, 1, 2]]_X$ code using the analogous protocol. (c) Fidelities averaged over groups of states that require the same number of logical operations. The filled bars correspond to the Bloch states for the original code and the hatched bars to those of the rotated version. On average, fidelities are higher for the rotated code, which is less sensitive to the dephasing. The error bars show standard deviations, determined as discussed in App.~\ref{tomo_details}.}
    \label{fig:bloch_inverted}
\end{figure*}

We use the $[[10, 1, 2]]_X$ code to span different states on the Bloch sphere, similar to Fig.~\ref{fig:bloch}, and compare its performance to the $[[10, 1, 2]]_Z$ code, which is shown in Fig~\ref{fig:bloch_inverted}. We now initially prepare $\ket{0}_L$ of the $[[10, 1, 2]]_X$ code and apply an $X$-rotation to create various states in the $YZ$-plane of the Bloch sphere (blue, green). Then, we switch to the $[[7, 1, 3]]$ code by measuring the three $Z$-stabilizers of the $[[7, 1, 3]]$ code and apply $\pi/2$ rotations about the $Y$- or $Z$-axis since these are not available transversally in the $[[10, 1, 2]]_X$ code. In doing so, we prepare eight additional states requiring non-Clifford gates (red) and two cardinal states (orange). By using this rotated $[[10, 1, 2]]$ code, fidelities improve on average by 0.046 for the states shown in blue and 0.027 for the green states.

This is due to the fact that the qubits spend less time in a decoherence-sensitive state and, during switching from $[[10, 1, 2]]$, $Z$-faults on data qubits cannot propagate to the auxiliary qubits and corrupt the switching syndrome. 
Therefore, it can be beneficial to adjust the theoretically constructed codes based on the knowledge of the setup's intrinsic error profile. 
However, for longer circuits with several mid-circuit measurements, as for the CNOT protocols shown in Fig.~\ref{fig:cnot}, the difference is less pronounced.

\end{document}